\documentclass[a4paper]{article}
\pdfoutput=1
\usepackage[round]{natbib}
\usepackage{amsmath,amsthm,amssymb}
\usepackage{float}
\usepackage{physics}
\usepackage{graphicx}
\usepackage{setspace}
\usepackage{caption}
\usepackage{subcaption}
\usepackage{fullpage}
\usepackage{authblk}
\usepackage{lineno}
\usepackage{adjustbox}
\usepackage{booktabs}
\usepackage{multirow}
\usepackage{bigstrut}
\usepackage{caption}
\usepackage{longtable}
\usepackage{pdflscape}
\usepackage{multicol}
\usepackage{hyperref}
\hypersetup{
    colorlinks=true,
    citecolor=blue,
}

\begin{document}

\title{Spatial and Temporal Evolution of Particle Migration in Gap-Graded Granular Soils: Insights from Experimental Observations}
\author[1]{V. S. Ramakrishna Annapareddy\footnote{Corresponding Author, email:
\href{v.annapareddy@uq.edu.au}{v.annapareddy@uq.edu.au}, ORCiD: 0000-0003-2816-3250}}
\author[1]{Adnan Sufian} 
\author[1]{Thierry Bore} 
\author[1]{Alexander Scheuermann} 
\affil[1]{School of Civil Engineering, The University of Queensland, Brisbane, Australia}
\date{}
\setcounter{Maxaffil}{0}
\renewcommand\Affilfont{\itshape\small}

\maketitle

\begin{abstract}
This study presents physical observations and insights into particle migration characteristics throughout the suffusion process. Using a purpose-built coaxial permeameter cell, suffusion experiments were conducted on idealised internally unstable gap-graded granular soils at varying fines content and hydraulic loading conditions. The specimens were prepared with a mixture layer comprising finer and coarser fractions underlying a coarse layer composed of the coarser fraction alone. This enabled the finer fraction within the mixture layer to migrate through the coarse layer with upward seepage flow. The local porosity profile along the specimen was determined using spatial time domain reflectometry and an inversion algorithm, which enabled the development of a novel field map of the difference in porosity from the initial condition. This field map provided a visual guide of the spatial and temporal variation in porosity and enabled particle migration internally within the specimen to be quantitatively characterised from onset to progression to washout. The limiting onset condition identified from the field map was shown to be comparable to that obtained using conventional approaches, thereby providing strong validation for the application of porosity-based field maps. As suffusion progressed, the height of infiltrating finer particles into the coarse layer increased linearly with time, while the overall rate of particle migration from the mixture layer to the coarse layer evolved in a non-linear manner with the rate of migration increasing as the specimen reached a complete mixture condition, where the finer fraction infiltrated the entire coarse layer. The attainment of a complete mixture condition was dependent on the fabric of the gap-graded soil. Specimens with an underfilled fabric showed a gradual migration process, while specimens with a transitionally underfilled fabric resulted in minimal particle migration followed by a very rapid formation of the complete mixture.
\end{abstract}

% \doublespacing
% \linenumbers
\setlength{\parindent}{0em}
\setlength{\parskip}{1em}

\section{Introduction}
Particle migration is a core feature of internal erosion failure in which seeping water dislodges and migrates finer particles from the body or foundation of embankment dams and levees. Suffusion is one mechanism through which internal erosion can initiate, and as particle migration progresses, it can ultimately lead to the breach of an embankment dam or levee. Suffusion occurs in internally unstable soils where the finer fraction of the soil is under low effective stress conditions \citep{shire2014fabric,langroudi2015comparison,indraratna2015,sufian2021influence}. The process of suffusion involves the migration of finer particles through the constrictions formed by the coarser particles with no significant change in the soil skeleton \citep{Kovacs1981,fannin2014distinct}. The migration of fines results in pore-scale changes in the soil structure, and consequently, local changes in the soil properties such as porosity and permeability \citep{nguyen2019experimental}. This study presents an experimental technique that can capture the spatial and temporal changes in local porosity and permeability throughout the entire suffusion process, enabling novel insights into particle migration characteristics internally within a gap-graded soil. This provides an improved understanding of the mechanisms that initiate and sustain the suffusion process.

Existing approaches for measuring the spatial or temporal variation in porosity in laboratory experiments include cumulative loss of fine particles \citep{ke2014triaxial,rochim2017effects}, visual observation of changes in layer heights and post-test sampling \citep{ke2012strength}, computed tomography scans \citep{homberg2012automatic,sufian2013microstructural,fonseca2014microstructural,nguyen2019experimental}, use of gamma rays and a scintillation counter \citep{alexis2004experimental,sibille2015description} and electromagnetic methods \citep{bore2018new,mishra2020characterisation} using conventional time domain reflectometry and spatial time domain reflectometry (spatial TDR) \citep{scheuermann2010fast,scheuermann2012determination,bittner2019determination}. Although these approaches have their advantages and limitations, spatial TDR is the only technique that provides measurements of the transient evolution of porosity profiles with sufficient accuracy and reasonable spatial and temporal resolution \citep{sufian2022physical}.

This study employs spatial TDR to present an in-depth investigation of the spatial and temporal variation in the local porosity profile resulting from the migration of particles throughout the suffusion process. This spatial and temporal information on local porosity and particle migration is prohibitive to obtain in existing experimental approaches, which generally measure eroded particles when they leave the specimen or infer locally after an experiment is complete by retrieving the specimen in layers and comparing pre and post-test particle size distributions \citep{kenney1985internal,moffat2006large,wan2008assessing,ke2012strength,benamar2019suffusion}. The main contribution of this study is the development of an experimental approach to quantitatively assess the rate of particle migration internally within a soil throughout the entire suffusion process at a local spatial and temporal resolution that has not been considered in existing studies. This is achieved by using a purpose-built coaxial permeameter \citep{bittner2019determination}, from which a novel difference in porosity field map was developed, allowing for particle migration characteristics to be quantified. Gap-graded granular soils with varying fines content are investigated under different hydraulic loading conditions. The local porosity obtained from spatial TDR is used to determine the amount, instantaneous and cumulative rates of particle migration within the test specimen throughout the suffusion process which would have not been possible with existing experimental approaches.    

\section{Suffusion experiments using coaxial permeameter}

\subsection{Experimental Apparatus}

\begin{figure}[ht]
    \centering
    \includegraphics[width=\textwidth]{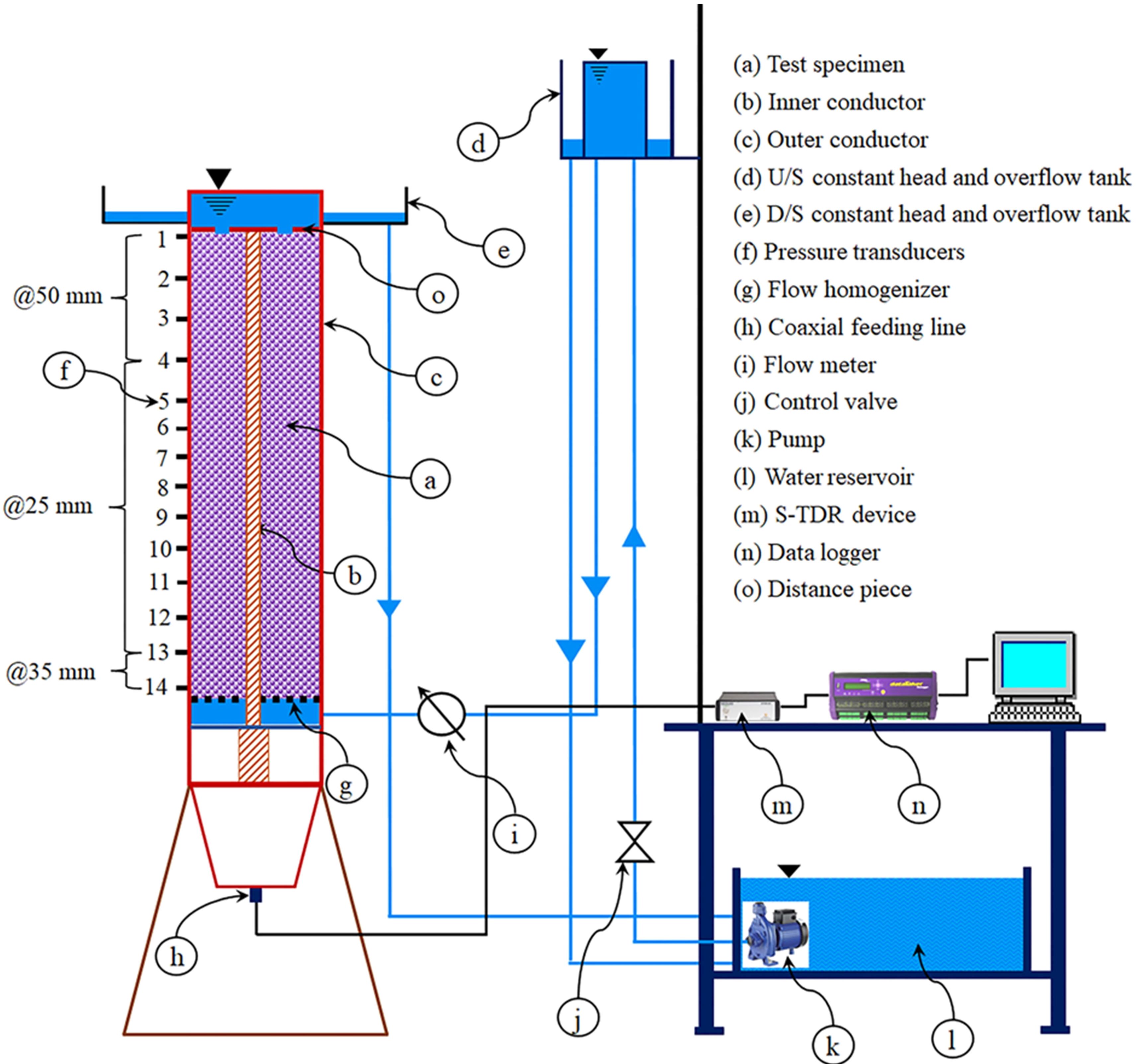}
    \caption{Schematic view of the experimental apparatus.}
    \label{fig:Schematic}
\end{figure}

The experimental setup resembles a laboratory constant head permeameter apparatus enhanced with capabilities to measure the spatial and temporal evolution of porosity using spatial TDR. The experimental setup consists of four main components: i) coaxial permeameter cell, ii) spatial TDR device, iii) hydraulic controlling unit, and iv) data acquisition system. A schematic of the experimental setup is provided in Fig.~\ref{fig:Schematic}. A detailed explanation of the design, calibration and validation of the experimental apparatus has been presented in \citet{bittner2019determination} and only a summary of the key features is presented, as this study focusses on the process of particle migration during suffusion experiments using the coaxial permeameter cell.

The copper-built permeameter cell is manufactured as a coaxial transmission line to facilitate electromagnetic measurements which enable the porosity profile along the height of the test specimen to be determined. The coaxial permeameter cell has an inner and an outer conductor as shown in Fig.~\ref{fig:Schematic} and the cell can accommodate a specimen of height up to 450 mm within the annulus formed by these two conductors. The outer conductor has an inner diameter of 151.9 mm, and the inner conductor has an outer diameter of 41.3 mm. A 40 mm wide and 420 mm high observation window on the outer wall of the cell allows visual inspection of particle migration during suffusion experiments.

A conventional TDR device measures the velocity of a high-frequency electromagnetic signal propagating through a waveguide that is usually inserted into the material under investigation. Within this study, the permeameter cell acts as the waveguide and the sample in the annulus is the material under investigation. The velocity of the electromagnetic signal depends on the permittivity (dielectric constant) of the sample filling the annulus. Conventional TDR is widely applied to monitor the moisture content in soils \citep{yu2004soil,cho2004spatial,jung2013new,curioni2018extending}, as the permittivity of the soil can be strongly correlated to the moisture content due to the dipole characteristic of water molecules. Conventional TDR has also been used to determine soil density \citep{drnevich2005time,jung2013new,bhuyan2018flat,curioni2018extending,bhuyan2021monitoring}. One of the limitations of conventional TDR is that it can only provide pointwise or averaged information rather than spatial information. The development of a fast inversion algorithm by \citet{schlaeger2005fast} enabled the computation of soil moisture profiles along the embedded probe or waveguide from the measured TDR trace. This approach is known as spatial TDR, and it has been extensively used for both field and laboratory scale applications which include obtaining moisture content distributions in levees, dikes \citep{scheuermann2009spatial}, and coastal dunes \citep{fan2015quantifying}, measuring pressure distribution \citep{scheuermann2008feasibility}, porosity profiles of water-saturated granular media \citep{scheuermann2012determination,bittner2019determination,sufian2022physical} and measuring moisture content in partially saturated soils \citep{mishra2018dielectric,yan2021application}. Therefore, a spatial TDR device is used in this study to obtain the electromagnetic measurements along the coaxial permeameter cell. From the measured TDR trace, the specimen’s porosity distribution along the cell could be obtained using an inversion algorithm which is detailed in \citet{bittner2019determination} and \citet{sufian2022physical}.

The hydraulic control system features a closed water cycle and applies upward seepage flow in the coaxial permeameter cell. Water enters the cell at its base from an upstream constant head overflow tank. To maintain a constant total head, a downstream constant head overflow tank is arranged at the top of the cell, which is connected to the downstream reservoir as shown in Fig.~\ref{fig:Schematic}. The difference in the water level between the upstream and downstream constant head overflow tank is indicative of the applied hydraulic head. The elevation of the upstream constant head overflow tank can be altered to apply different hydraulic loading conditions. 38 mm diameter hoses are used for all connections to minimise head loss. To impose a relatively uniform flow boundary condition at the upstream side of the specimen, a 10 mm thick perforated plate made of PMMA (polymethyl methacrylate) was placed just above the flow inlet, which acted as a flow homogeniser. Another advantage of using PMMA is that the relatively low dielectric permittivity of PMMA results in a clear peak in the TDR trace, which signifies the start of the specimen. At the end of the inner conductor, a metallic short-circuit is positioned, which results in a sharp drop in the TDR trace, allowing for the end of the cell to be easily determined. The travel time of the voltage step signal can be defined using the impedance mismatch at the start and end of the cell, which is a key step in the TDR analysis to determine the height of the specimen. 

The pore-water pressure within the specimen is monitored along the sidewall of the cell using 14 pressure transducers (PT) which were arranged as shown in Fig.~\ref{fig:Schematic}. PT14 and PT13 are located at 10 mm and 45 mm above the base of the observation window, respectively. The pressure transducers from PT13 to PT4 are spaced at an interval of 25 mm, while the spacing is kept at 50 mm from PT4 to PT1. These point measurements of pore water pressure are used to calculate the hydraulic gradient within the test specimen. Note that the endpoints of the pressure transducers are flush with the inner wall of the cell. This may cause some inaccuracies in the measurement of the pore water pressures particularly if there are preferential flow paths along the circumference of the cell. The volumetric flow rate is measured using a flowmeter located upstream of the cell to avoid the blockage of fine particles within the flowmeter.

\subsection{Materials, specimen configuration and preparation}

Suffusion experiments in the coaxial permeameter cell were conducted with an idealised granular soil comprising soda-lime glass beads with a specific gravity of 2.5. Previous studies have highlighted the suitability and benefits of using glass beads as a substitute for real soil \citep{tomlinson2000seepage,scheuermann2012determination,zakaraya2015discussion,bittner2017experimental,sufian2022physical}. Glass beads have a constant dielectric permittivity independent of its particle size which is an additional advantage from a spatial TDR viewpoint. Moreover, by using glass beads, the influence of differences in the mineralogy of real soil particles can be avoided. Despite its high roundness, glass beads provided a fundamental basis to better understand the phenomenon of particle migration during suffusion experiments.

 \begin{figure}[ht]
     \centering
     \includegraphics[width=0.6\textwidth]{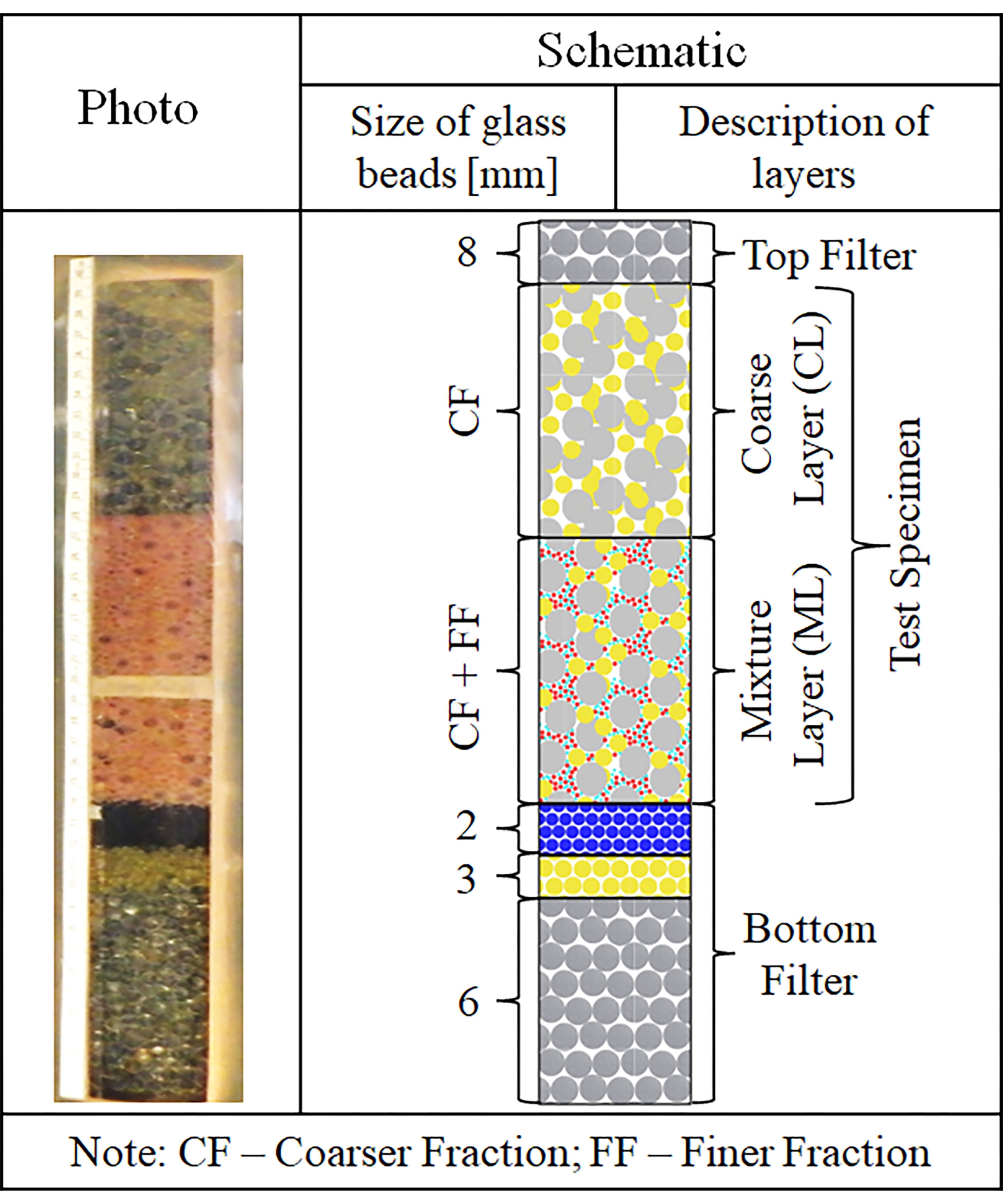}
     \caption{Photo and schematic of the specimen configuration}
     \label{fig:SpecimenConfig}
 \end{figure}

The specimen configuration is graphically illustrated in Fig.~\ref{fig:SpecimenConfig} and comprises four zones: a bottom filter, a mixture layer, a coarse layer, and a top filter. The bottom filter consists of approximately 10 cm layer of 6 mm diameter glass beads, with overlying layers of 3 mm and 2 mm diameter glass beads, each approximately 2 cm in height. The primary function of the bottom filter is to prevent significant downward loss of fine particles through the flow homogeniser, which is further verified in the subsequent analysis. The mixture layer (ML) consists of a combination of the coarser fraction and the finer fraction and is approximately 13.5 cm in height. The coarser fraction comprises 3 mm and 8 mm diameter monodisperse glass beads at a ratio of 1:3 resulting in a stepped particle size distribution in Fig. 3, while the finer fraction comprises particles ranging from 0.3 to 0.6 mm. The coarser and finer fractions are mixed in different proportions to obtain different particle size distributions as discussed below. The coarse layer (CL) overlying the mixture layer is approximately 13.5 cm in height and consists of the coarser fraction alone. The top filter consists of an approximately 3 cm layer of 8 mm diameter glass beads. The test specimen (ML and CL) was intentionally prepared in this fashion to enable spatial TDR to capture the changes in porosity caused by the migration of the finer fraction from the ML to the CL during suffusion experiments. The subsequent analysis demonstrates that the test specimen was sized appropriately to ensure that the entire suffusion process could be investigated.
 
As the bottom and top filters are monodisperse packings, they were prepared under dry conditions. While the test specimen (ML and CL) was prepared using the moist tamping method to minimise the segregation of different-sized glass beads upon placement into the coaxial permeameter cell \citep{ladd1978preparing}.  In this method, approximately 4-6\% of moisture content by dry mass of glass beads is added to the mixture and mixed thoroughly. The specimen was prepared in 10 layers, with each layer approximately 2.6 cm in height, to ensure that the specimen was as homogenous as possible. The homogeneity of the test specimen was verified by the local porosity profile obtained from spatial TDR data in the subsequent analysis. Once the specimen was prepared, it was saturated gradually by upward seepage at an applied head of $<$ 1 cm, resulting in a very low flow rate which did not lead to observable particle migration prior to the onset of suffusion, as verified in the subsequent analysis.

\subsection{Experimental program}

\begin{figure}[ht]
     \centering
     \includegraphics[width=0.6\textwidth]{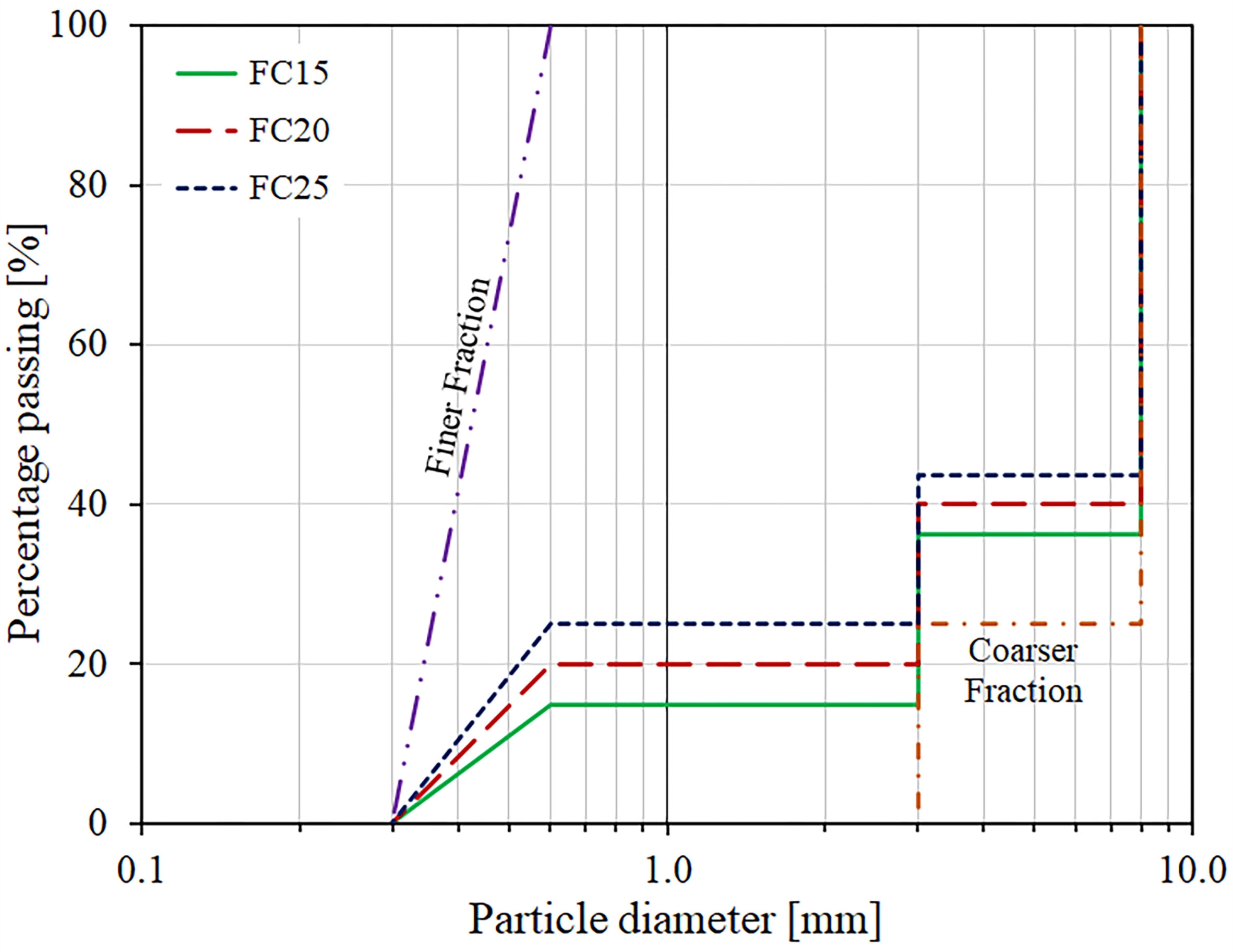}
     \caption{PSDs of the materials investigated in this study}
     \label{fig:PSDs}
 \end{figure}

Three different samples were prepared by mixing the coarser and finer fractions with 15\%, 20\%, and 25\% fines content, (termed FC15, FC20 and FC25 in Fig.~\ref{fig:PSDs}) forming gap graded particle size distributions (PSDs) with a gap ranging the particle sizes from 0.6 to 3.0 mm. The PSDs of the coarser and finer fractions are also shown in Fig.~\ref{fig:PSDs}. \citet{skempton1994experiments} distinguished the soil fabric based on fines content ($f_c$) with \emph{underfilled fabric} for $f_c \leq 24\%$, \emph{transitional fabric} for $24\% < f_c < 35\%$, and \emph{overfilled fabric} for  $f_c \geq 35\%$. A maximum fines content of 25\% is considered in this study to ensure that the test specimen had an underfilled or transitionally underfilled fabric, where the finer fraction resided predominantly in the pore space of the coarser fraction, which formed the load-bearing soil skeleton. This assumption of underfilled or transitionally underfilled fabric was valid as no measurable settlements were observed in all experiments. Note that settlements can be quantified from spatial TDR data as demonstrated in the filtration experiments presented in \citet{sufian2022physical}.
 
\begin{figure}[ht]
     \centering
     \includegraphics[width=0.6\textwidth]{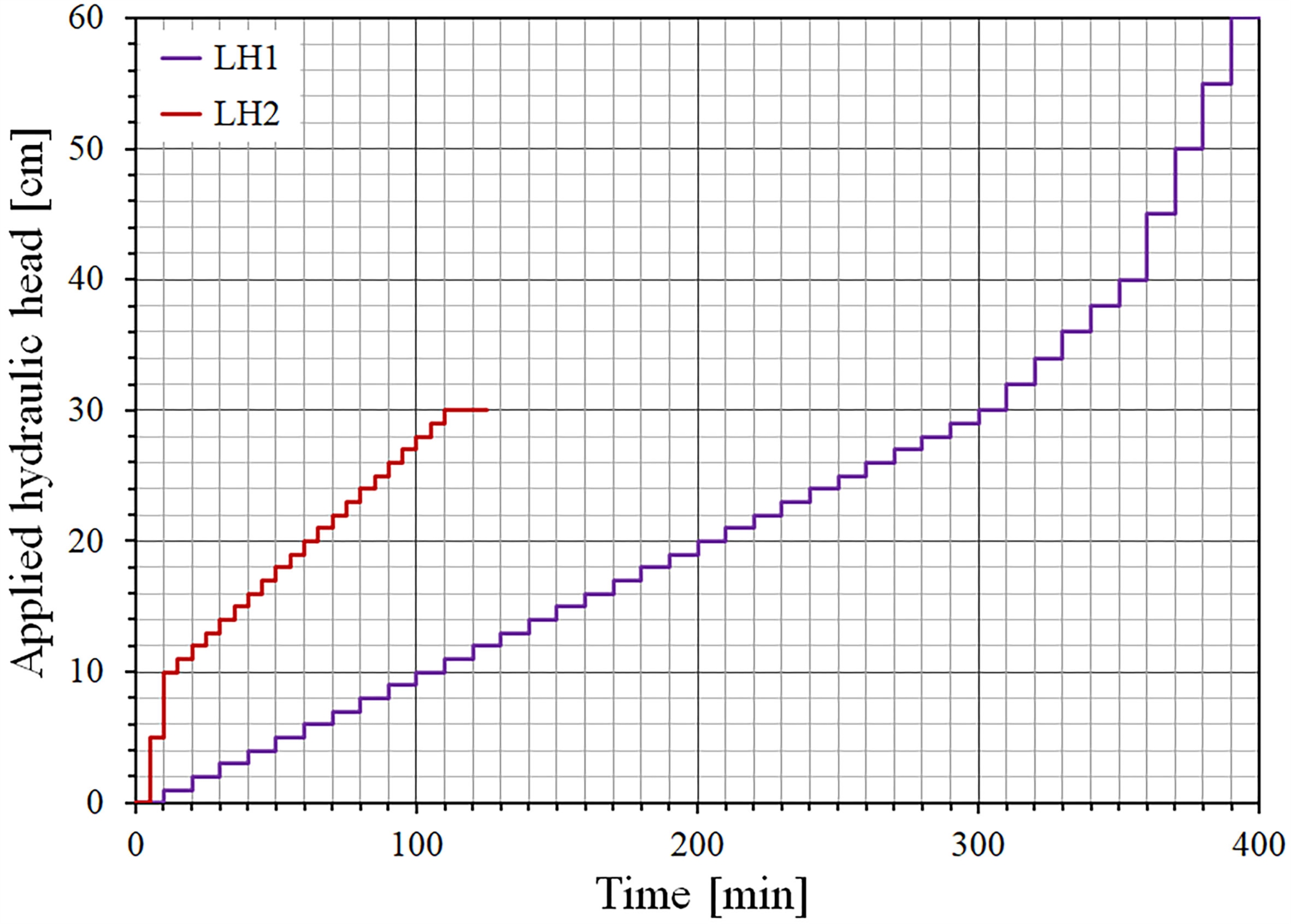}
     \caption{Hydraulic loading conditions applied in this study}
     \label{fig:Loading}
 \end{figure}

The three PSDs were tested under two different hydraulic loading conditions termed LH1 and LH2 as shown in Fig.~\ref{fig:Loading}. In LH1, the hydraulic head was increased by 1 cm every 10 minutes until an applied head of 30 cm, thereafter, the head was increased in steps of 2 cm and 5 cm every 10 minutes for the head ranges from 30 to 40 cm and 40 to 60 cm, respectively. In LH2, the hydraulic head between 0 and 10 cm was increased in steps of 5 cm every 5 minutes, after which the head was increased in steps of 1 cm every 5 minutes up to an applied head of 30 cm. Therefore, the rate of hydraulic loading in LH2 is higher compared to LH1.

Table~\ref{tab:Table4.1} shows the complete list of experiments performed in this study. The nomenclature for each test is $FCx\_LHy$, where \textit{x} can take the value of 15 or 20 or 25 to indicate the fines content in the ML and \textit{y} can take a value of 1 or 2 to specify which hydraulic loading condition is considered. Furthermore, Table~\ref{tab:Table4.1} lists the dry density ($\rho_d$), and initial height of the mixture layer ($h_{ML}$) and coarse layer ($h_{CL}$) at the end of specimen preparation, along with the initial average porosity of the mixture layer ($n_{avg}^{ML}$) and coarse layer ($n_{avg}^{CL}$) which were determined based on the dry mass of glass beads and the layer heights. The CL has approximately the same porosity in all the test cases, while the ML has reduced porosity and increased dry density with increasing fines content as the fines are filling the void space of the coarse particles. 

\begin{table}[ht]
\caption{Experimental program using the coaxial permeameter cell}
\begin{center}
\begin{tabular}{@{} lp{1.2cm}p{3.0cm}p{1.4cm}p{1.4cm}p{1.4cm}p{1.2cm}p{1.2cm} @{}}
\toprule
Test ID & $f_c$ [\%] & Hydraulic loading condition & $\rho_d$ [g/cm$^3$] & $h_{ML}$ [mm] & $h_{CL}$ [mm] & $n_{avg}^{ML}$ [--] & $n_{avg}^{CL}$ [--] \\
\midrule
      FC15\_LH1&15&LH1$^\ast$&1.86&133&135&0.25&0.36 \\
      FC20\_LH1&20&LH1&1.94&134&134&0.22&0.34 \\
      FC25\_LH1&25&LH1&2.07&136&135&0.17&0.35 \\
      FC15\_LH2&15&LH2&1.86&133&134&0.25&0.34 \\
      FC20\_LH2&20&LH2&1.91&135&133&0.23&0.33 \\
      FC25\_LH2&25&LH2&2.04&136&133&0.18&0.33 \\
\bottomrule
\end{tabular}
\end{center}
    $^\ast$ The maximum applied hydraulic head in FC15\_LH1 is 50 cm rather than 60 cm, as per hydraulic loading condition 1
\label{tab:Table4.1}
\end{table}

\subsection{Geometric assessment of internal instability}

The susceptibility to suffusion is dependent on the fines content and gap ratio of the gap-graded soil for underfilled fabric as well as relative density for transitionally underfilled fabric \citep{shire2014fabric,ahmadi2020modelling}. As discussed above, a maximum fines content of 25\% was considered to ensure that the gap-graded soils consisted of underfilled or transitionally underfilled fabric, such that particle migration would not lead to significant changes in the soil skeleton. The PSDs in Fig.~\ref{fig:PSDs} indicate that particle sizes ranging from 0.6 to 3.0 mm are absent and the gap ratio of 5 was chosen to ensure that the test specimen was internally unstable. This was advantageous as it enabled particle migration to be explored across the complete suffusion process. To confirm the internal instability of the PSDs in Fig.~\ref{fig:PSDs}, numerous geometric criteria proposed in prior studies were considered \citep{istomina1957filtration, kezdi1979, Kovacs1981, kenney1985internal, burenkova1993, Liu2005, indraratna2007, li2008comparison,  chang2013,li2013capillary,indraratna2015,to2018quick}. The test specimen was assessed against these criteria using the geometric properties listed in Table~\ref{tab:Table4.2}. Most of the properties in Table~\ref{tab:Table4.2} are readily obtained from the PSD. The average porosity ($n_{avg,CF}^{ML}$) of the coarser fraction in the ML is provided in Table~\ref{tab:Table4.2} and can be obtained from $n_{avg}^{ML}$ and $f_c$ as shown below:
\begin{equation}
n_{avg,CF}^{ML} = \frac{V-V_{CF}}{V}=\frac{V-V_S}{V}+\frac{f_c}{100}\frac{V_S}{V}=n_{avg}^{ML}+\frac{f_c}{100}\left( 1-n_{avg}^{ML} \right)
\label{eq:1}
\end{equation}
where $V$ is the total volume of the ML, $V_S$ is the total solid volume of the ML,$V_{CF} = V_S-V_{FF}$ is the solid volume of the coarser fraction in the ML, $V_{FF}$ is the solid volume of the finer fraction, and $f_c$ is the fines content which is defined as $\left ( V_{FF}/V_S \right)\times 100$ . The average pore diameter of the coarser fraction in the ML ($O_{50,CF}^{ML}$) was determined using the model proposed by \citet{Kovacs1981}, which is a function of $n_{avg,CF}^{ML}$ and the mean particle diameter of the coarser fraction, $d_e^{CF}$:
\begin{equation}
O_{50,CF}^{ML} = 4 \left(\frac{n_{avg,CF}^{ML}}{1-n_{avg,CF}^{ML}}\right) \left(\frac{d_e^{CF}}{\alpha_D}\right)
\label{eq:2}
\end{equation}							                 
and 
\begin{equation}
d_e^{CF} = \frac{1}{\Sigma_i \left(\Delta f_i^{CF}/d_i^{CF}\right)}
\label{eq:3}
\end{equation}								           
where $\Delta f_i^{CF}$ and $d_i^{CF}$ are the weight and average diameter of particles in the $i^{th}$ bin of the PSD curve, respectively, and $\alpha_D$ is the shape factor which is approximately equal to 6 for spherical particles.

\begin{table}[ht]
\begin{center} 
\caption{Experimental program using the coaxial permeameter cell} \label{tab:Table4.2}%
\begin{adjustbox}{width=1\textwidth} 
\begin{tabular}{@{} lp{0.7cm}p{0.7cm}p{0.8cm}p{0.8cm}p{0.7cm}p{0.8cm}p{0.8cm}p{1.3cm}p{0.6cm}p{0.7cm}p{0.6cm}p{1cm}p{1cm} @{}}
\toprule
PSD & $d_{10}$ [mm] & $C_u$ [--]& $d_{15}^{CF}$ [mm] & $d_{85}^{FF}$ [mm] & $d_{e}$ [mm] & $d_{e}^{CF}$ [mm] & $d_{e}^{FF}$ [mm] & $(H/F)_{min}$ [--] & $h'$ [--] & $h''$ [--] & $G_r$ [--] & $n_{avg,CF}^{ML}$ [--] & $O_{50,CF}^{ML}$ [mm]\\
\midrule
FC15&0.48&16.67&3.00&0.54&1.98&5.65&0.41&0.00&1.00&13.33&5.00&0.36&2.14 \\
FC20&0.42&18.82&3.00&0.54&1.63&5.65&0.41&0.00&1.00&15.69&5.00&0.38&2.31 \\
FC25&0.39&20.25&3.00&0.54&1.39&5.65&0.41&0.00&1.00&17.78&5.00&0.38&2.31 \\
\bottomrule
\end{tabular}
\end{adjustbox} 
\end{center} 
   Note: $d_{10}=$ particle diameter at 10\% mass passing in the finer fraction; $C_u=$ coefficient of uniformity; $d_{15}^{CF}=$ particle diameter corresponding to the 15\% mass passing in the coarser fraction; $d_{85}^{FF}=$ particle diameter corresponding to the 85\% mass passing in the finer fraction; $d_{e}=$ mean particle diameter; $d_{e}^{CF}=$ mean particle diameter of coarser fraction (Eq.~\ref{eq:3}); $d_{e}^{FF}=$ mean particle diameter of finer fraction; $F=$ proportion of mass at any particle diameter, $d$; $H=$ proportion of mass between the particle diameter $d$ and $4d$; $h'=d_{90}/d_{60}$ and $h''= d_{90}/d_{15}$ are conditional uniformity factors; $G_r=$ gap ratio; $n_{avg,CF}^{ML}=$ average porosity of coarser fraction the mixture layer (Eq.~\ref{eq:1}); $O_{50,CF}^{ML}=$ average pore diameter of the coarser fraction in the mixture layer (Eq.~\ref{eq:2})
\end{table}%

Table~\ref{tab:Table4.3} demonstrates that the vast majority of these geometric criteria identified the ML to be internally unstable, and hence, susceptible to suffusion. The experimental observations in this study also indicated that the test specimen in all the experiments was internally unstable.

\begin{table}
\caption{Summary on the geometric assessment of internal instability of the test specimens in this study}
\begin{center}
%\begin{adjustbox}{width=1\textwidth} 
\begin{tabular}{@{} lp{1.4cm}p{1cm}p{1.2cm}p{1.5cm}p{1.5cm}p{0.9cm}p{1.4cm}p{1.2cm}p{1cm} @{}}
\toprule
Test ID & \citet{istomina1957filtration} & \citet{kezdi1979} & \citet{Kovacs1981} & \citet{kenney1985internal} & \citet{burenkova1993} & \citet{Liu2005} & \citet{li2008comparison} & \citet{chang2013} & This Study \\
\midrule
      FC15\_LH1&B&U&U&U&U&U&U&U&U \\
      FC20\_LH1&B&U&U&U&U&U&U&S&U \\
      FC25\_LH1&U&U&U&U&U&B&U&S&U \\
      FC15\_LH2&B&U&U&U&U&U&U&U&U \\
      FC20\_LH2&B&U&U&U&U&U&U&S&U \\
      FC25\_LH2&U&U&U&U&U&B&U&S&U \\
\bottomrule
\end{tabular}
%\end{adjustbox}
\end{center}
    Note: S $=$ stable; B $=$ borderline; and U $=$ unstable.
\label{tab:Table4.3}
\end{table}

\section{Experimental results}

\begin{figure}
     \centering
     \includegraphics[width=0.9\textwidth]{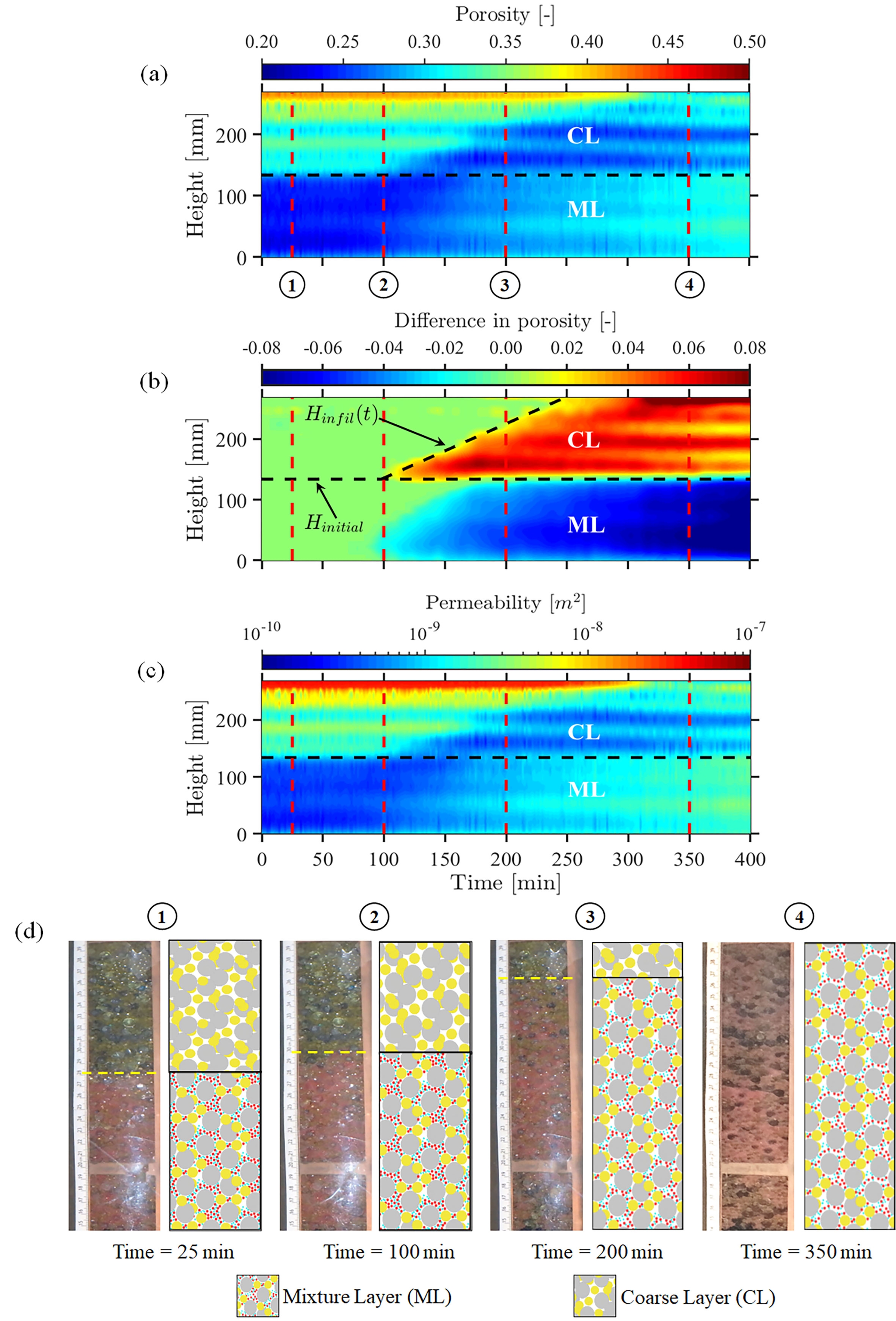}
     \caption{Test case of FC20\_LH1 (a) Porosity field map (b) Field map of difference in porosity, with lines of $H_{infil} (t)$ and $H_{initial}$ (c) Permeability field map (d) Visual and schematic representation of particle migration during suffusion experiments. The image of coaxial permeameter cell at different times illustrates the development of complete mixture condition resulting from particle migration}
     \label{fig:TypicalResult}
 \end{figure}

The spatial and temporal variation of porosity during the suffusion experiments can be obtained from spatial TDR data and analysed with different approaches. The transient evolution of porosity distribution along the height of the sample is graphically presented in the porosity field map shown in Fig.~\ref{fig:TypicalResult}a, where the colour indicates local porosity at a given specimen height and time. Fig.~\ref{fig:TypicalResult}b shows the field map of the difference in porosity, which is obtained by considering the changes in local porosity from the initial porosity profile, where the colour signifies the loss (red colour) or gain (blue colour) in porosity. As permeability is a function of porosity, the spatial and temporal evolution of permeability can also be obtained from the measurements of local porosity using the modified Kozeny-Carman model ($K_{local,TDR}^{K-C}$) proposed by \citet{annapareddy2022computation}:
\begin{equation}
    K_{local,TDR}^{K-C}=\frac{n_{local}^3 d_{eff}^2}{C \left(1-n_{local}\right)^2}
\label{eq:4}
\end{equation}
where   is the local porosity obtained from spatial TDR data, C is the shape factor which can be approximated as 180 for spherical particles (a reasonable assumption for glass beads) and  is the effective particle diameter, which is given by:
\begin{equation}
    d_{eff}^{-1}=\psi_{CF} \left(d_e^{CF}\right)^{-1} + \psi_{FF} \left(d_e^{FF}\right)^{-1}
\label{eq:5}
\end{equation}
where $d_e^{CF}$ and $d_e^{FF}$ are the mean particle diameters of the coarser and finer fractions, respectively, and can be obtained from their respective PSDs given in Fig.~\ref{fig:PSDs}. $\psi_{CF}$ and $\psi_{FF}$ are the volume proportions of the coarser and finer fractions, respectively, and are calculated based on the porosity of the specimen ($n_{local}$), along with the porosity of the coarser and finer fractions, as detailed in \citet{annapareddy2022computation}. The permeability field map is displayed in Fig.~\ref{fig:TypicalResult}c, where the colour indicates permeability at a given height and time during the suffusion experiment.

Four characteristic conditions of a typical suffusion experiment are identified in the field maps in Fig.~\ref{fig:TypicalResult}: (1) distinct and stable transition from ML to CL; (2) onset of particle migration; (3) progression of suffusion with the migration of fines from the ML to the CL; and (4) complete migration of the finer fraction through the CL. These four characteristic conditions are schematically illustrated in Fig.~\ref{fig:TypicalResult}d and labelled 1 to 4. At the beginning of the experiment, the porosity in the ML and CL are distinct, and as expected, the initial porosity and permeability in the ML are significantly lower than in the CL [Fig.~\ref{fig:TypicalResult}a and \ref{fig:TypicalResult}c].

The porosity and permeability of the ML and CL do not vary significantly until the specimen reaches the limiting onset condition, where the finer fraction in the ML experience local fluidisation. This stage indicates the onset of the suffusion process. Following the onset, any increment in hydraulic gradient results in the migration of the finer fraction from the ML to CL which signifies the progression of suffusion. The finer fraction from the ML migrates into the voids of CL, which results in a local increase and reduction in the porosity of ML and CL, respectively [Fig.~\ref{fig:TypicalResult}a$-$c]. While the progression phase can be investigated at a constant applied head, this study investigated the progression of suffusion with successive increments in the applied head, as indicated in Fig.~\ref{fig:Loading}. Therefore, the progression phase is analysed both in terms of temporal evolution as well as the influence of changes in local hydraulic gradient in the subsequent analysis. Another important characteristic of the progression of suffusion is that the infiltration height of the finer fraction into the CL gradually increases with an increase in the applied hydraulic head leading to a complete mixture condition, which is defined by the condition where the finer fraction has infiltrated the entire CL as illustrated in Fig.~\ref{fig:TypicalResult}d.

Although the onset of suffusion can be inferred from any of the field maps presented in Fig.~\ref{fig:TypicalResult}, it was found that both onset and progression of suffusion could be better characterised using the field map based on the difference in porosity [Fig.~\ref{fig:TypicalResult}b], which is used in the subsequent analysis. The infiltration height of the finer fraction into the CL, $H_{infil} (t)$, at a given time, can be readily identified in the field map in Fig.~\ref{fig:TypicalResult}b. The intersection of $H_{infil} (t)$ and the line representing the initial height of the interface between the ML and the CL, $H_{initial}$, provides the basis for the onset of suffusion, while the gradient of $H_{infil} (t)$ line is a measure of the rate of infiltration, which is approximated as the rate of progression. However, it should be noted that this is just an approximation of the rate at which the finer fraction has infiltrated into new pore spaces of the CL for the first time. While there is no increase in the infiltration height at a given applied hydraulic head, the migration of finer fraction may still be occurring within the infiltrated height, and this is further explored in the sub-section 4.5.1. Although the finer fraction infiltrates into the CL with the progression of suffusion, the terminology ML and CL in the subsequent analysis refers to the initial delineation of the mixture and coarse layer upon sample preparation.

\clearpage

\section{Onset of suffusion}

\subsection{Based on difference in porosity field map from spatial TDR}

The onset of particle migration occurs when the driving hydrodynamic forces due to upward seepage on the finer particles exceed the resisting gravitational and inter-particle contact forces. At this limiting condition, the finer particles are locally fluidised and will subsequently migrate through the pore space of the coarser fraction. The limiting onset condition can be inferred from the difference in porosity field map by the intersection of the $H_{infil} (t)$ and $H_{initial}$ lines, and the time at which this occurs ($t_{onset}$) can also be determined. The difference in porosity field maps for all cases in Table~\ref{tab:Table4.1} is shown in Figs.~\ref{fig:FC15LH1}$-$\ref{fig:FC25LH2}. The onset of suffusion can be observed for both hydraulic loading conditions (LH1 and LH2). Figs.~\ref{fig:FC15LH1}$-$\ref{fig:FC25LH2} also show the applied hydraulic head, measured flow rate, average hydraulic gradient ($i_{avg}$) across the mixture layer ($i_{avg}^{ML}$), coarse layer ($i_{avg}^{CL}$) and the test specimen ($i_{avg}^{ML+CL}$), and the average porosity ($n_{avg}$) across the mixture ($n_{avg}^{ML}$) and coarse ($n_{avg}^{CL}$) layers. The average hydraulic gradients ($i_{avg}^{ML}$, $i_{avg}^{CL}$ and $i_{avg}^{ML+CL}$) are calculated from the measured pore water pressures along the cell using pressure transducers. $n_{avg}^{ML}$ and $n_{avg}^{CL}$ are computed by averaging the local porosity profile obtained from spatial TDR data. 

The dashed vertical line in Figs.~\ref{fig:FC15LH1}$-$\ref{fig:FC25LH2} indicate $t_{onset}$ and is listed in Table~\ref{tab:Table4.4}. The flow rate corresponding to the onset of suffusion, $Q_{onset}$, is also listed in Table~\ref{tab:Table4.4}, which can be used to determine the critical seepage velocity by considering the cross-sectional area of the specimen and the average porosity of the mixture layer. $Q_{onset}$ was significantly influenced by fines content in the ML with $Q_{onset}$ decreasing with increasing $f_c$. An increase in $f_c$ reduced the pore volume, as finer particles filled the void space of the ML, which in turn caused a reduction in flow rate. In addition, a slight dependence on the rate of hydraulic loading was observed. The increased rate of loading in LH2 resulted in a slightly higher $Q_{onset}$, with the only exception being FC15\_LH2.

The hydraulic gradient in the mixture layer at the onset of suffusion, $i_{onset}^{TDR}$, is also listed in Table~\ref{tab:Table4.4}. $i_{onset}^{TDR}$ showed a strong dependence on the fines content in the ML with a higher $f_c$ resulting in a higher $i_{onset}^{TDR}$. The rate of hydraulic loading showed only a slight influence on $i_{onset}^{TDR}$, with lower $i_{onset}^{TDR}$ noted for LH1. The influence of the rate of hydraulic loading was more pronounced in specimens with an underfilled fabric ($f_c=15\%$ and 20\%) compared to the specimens with a transitionally underfilled fabric ($f_c=25\%$). Leading up to the limiting onset condition, $i_{avg}^{ML}$ is significantly higher than $i_{avg}^{CL}$, while the $n_{avg}^{ML}$ is considerably lower than $n_{avg}^{CL}$. This is expected as the entire finer fraction existed only in the ML, leading to lower porosity and increased head loss in the ML prior to the onset condition. 

\begin{table}[ht]
\caption{Key characteristics at limiting onset and complete mixture conditions} \label{tab:Table4.4}%
%\footnotesize
\centering
\begin{adjustbox}{width=\textwidth}
\begin{tabular}{@{} lp{1cm}p{1cm}p{1cm}p{1cm}p{1cm}p{1cm}p{1cm}p{1.5cm} p{1cm}p{1cm}p{1cm}p{1.8cm} @{}}
\toprule
Test ID 
& \multicolumn{8}{c}{limiting onset condition} 
& \multicolumn{4}{c}{complete mixture condition} \\
\cmidrule(lr){2-9} \cmidrule(lr){10-13}
& $t_{onset}$ [min] & $Q_{onset}$ [L/min] & $i_{onset}^{TDR}$ [--] & $i_{onset}^{Darcy}$ [--] & $i_{crit}^{Terz}$ [--] & $E_{onset}$ [J] & $k_{avg}^{Darcy}$ [cm/s] & $k_{avg}^{Chapuis}$ [cm/s] & $t_{CM}$ [min]  & $Q_{CM}$ [L/min] & $i_{CM}$ [--] & $\Dot{H}_{infil}$ [mm/min] \\
\midrule
FC15\_LH1&135&5.26&0.19&0.15&1.12&10.36&0.17&0.05&315&10.40&0.30&0.75 \\
FC20\_LH1&98&2.37&0.32&0.29&1.16&3.57&0.04&0.03&248&07.33&0.45&0.89 \\
FC25\_LH1&155&0.91&0.78&0.83&1.24&4.41&0.01&0.01&338&07.23&1.41&0.14 \\
FC15\_LH2&15&4.61&0.25&0.20&1.12&21.73&0.12&0.05&–-&–-&--&0.97 \\
FC20\_LH2&12&3.52&0.38&0.33&1.15&10.45&0.05&0.03&123&09.13&0.57&0.78 \\
FC25\_LH2&38&1.08&0.79&0.76&1.22&60.71&0.01&0.01&100&05.72&1.00&0.47 \\
\bottomrule
\end{tabular}
\end{adjustbox}
\end{table}%

\begin{landscape}

\pagestyle{empty}

\newenvironment{Figure}
  {\par\medskip\noindent\minipage{\linewidth}}
  {\endminipage\par\medskip}

\begin{multicols}{3}

\begin{Figure}
     \centering
     \captionsetup{type=figure,justification=centering}
     \includegraphics[width=0.92\textwidth]{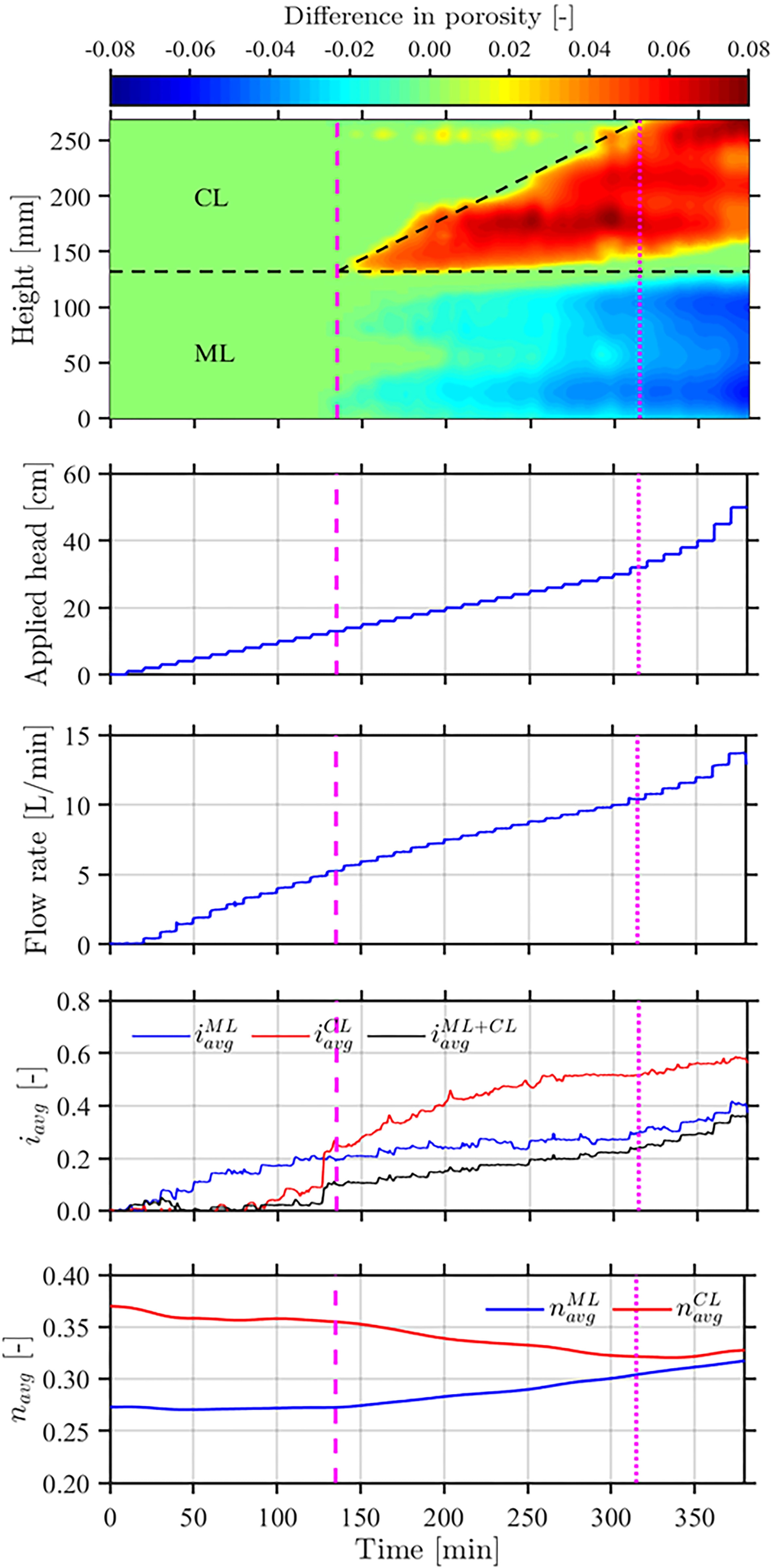}
     \captionof{figure}{Experimental observations for FC15\_LH1 case. In Figs.~\ref{fig:FC15LH1}$-$\ref{fig:FC25LH2}, the limiting onset condition is indicated by a dashed line and the complete mixture condition is indicated by a dotted line}
     \label{fig:FC15LH1}
\end{Figure}
\begin{Figure}
     \centering
     \captionsetup{type=figure,justification=centering}
     \includegraphics[width=0.92\textwidth]{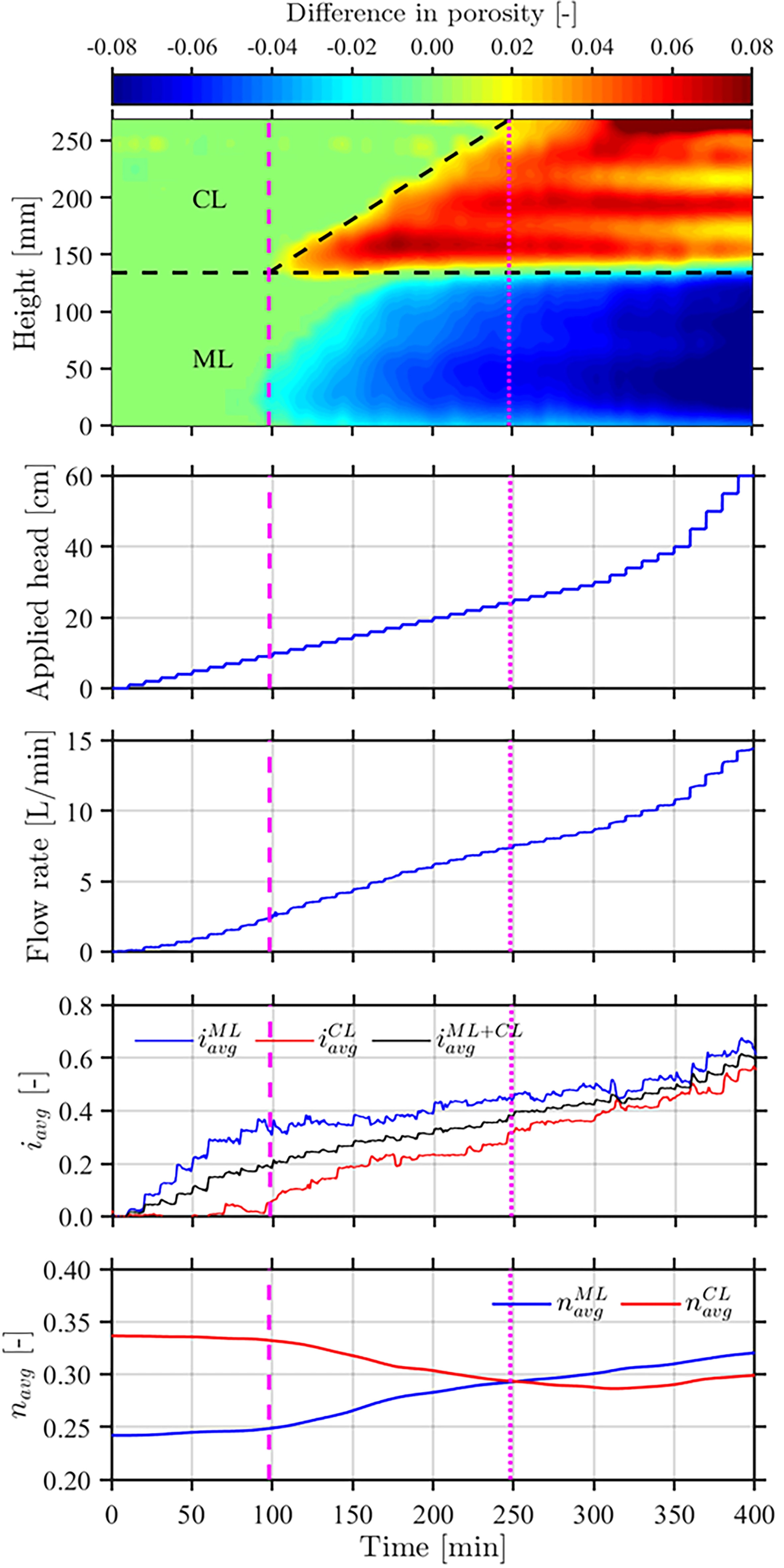}
     \captionof{figure}{Experimental observations for FC20\_LH1 case}
     \label{fig:FC20LH1}
\end{Figure}
\begin{Figure}
     \centering
     \captionsetup{type=figure,justification=centering}
     \includegraphics[width=0.92\textwidth]{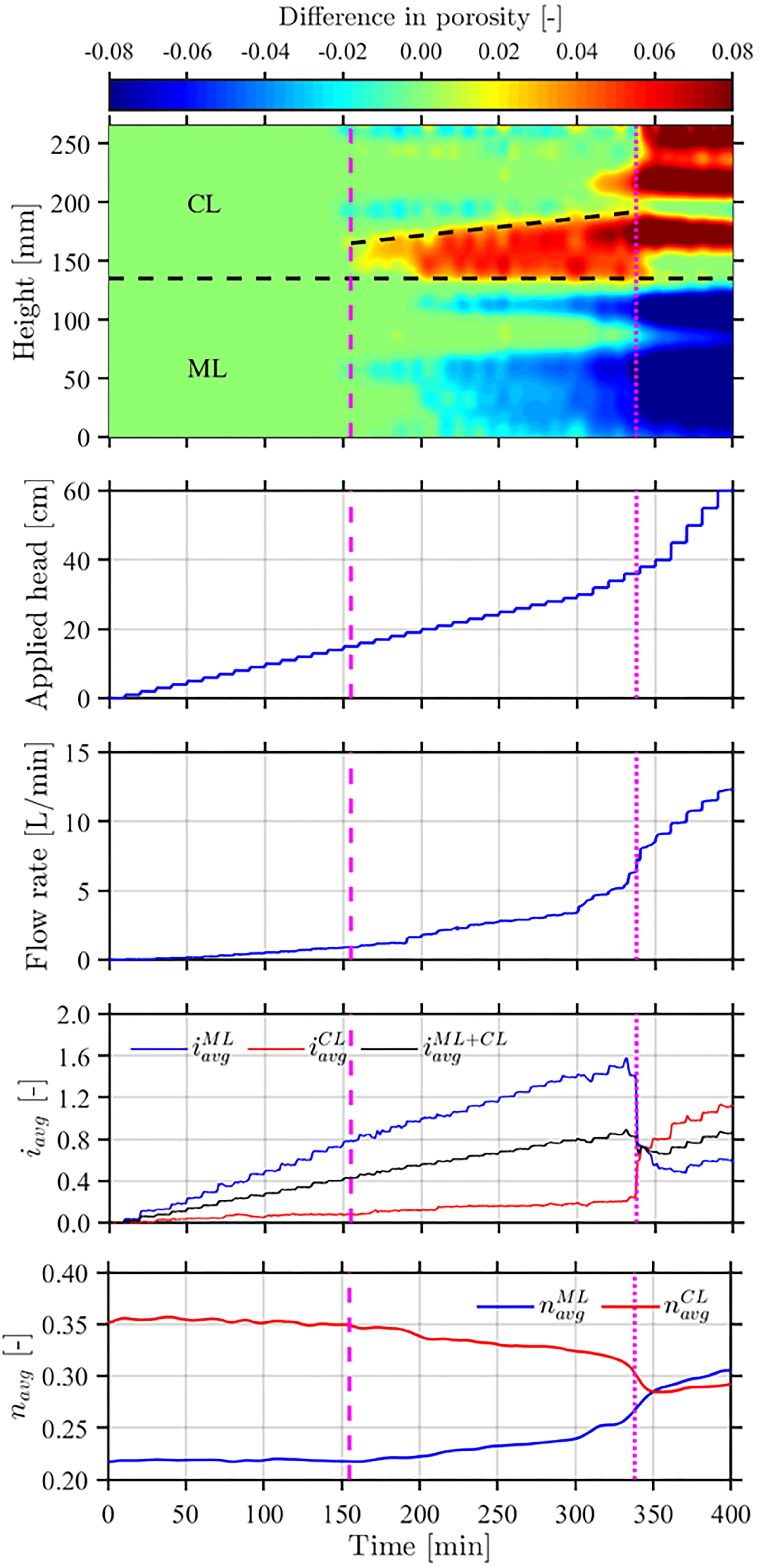}
     \captionof{figure}{Experimental observations for FC25\_LH1 case.}
     \label{fig:FC25LH1}
\end{Figure}
 
\begin{Figure}
     \centering
     \captionsetup{type=figure,justification=centering}
     \includegraphics[width=0.92\textwidth]{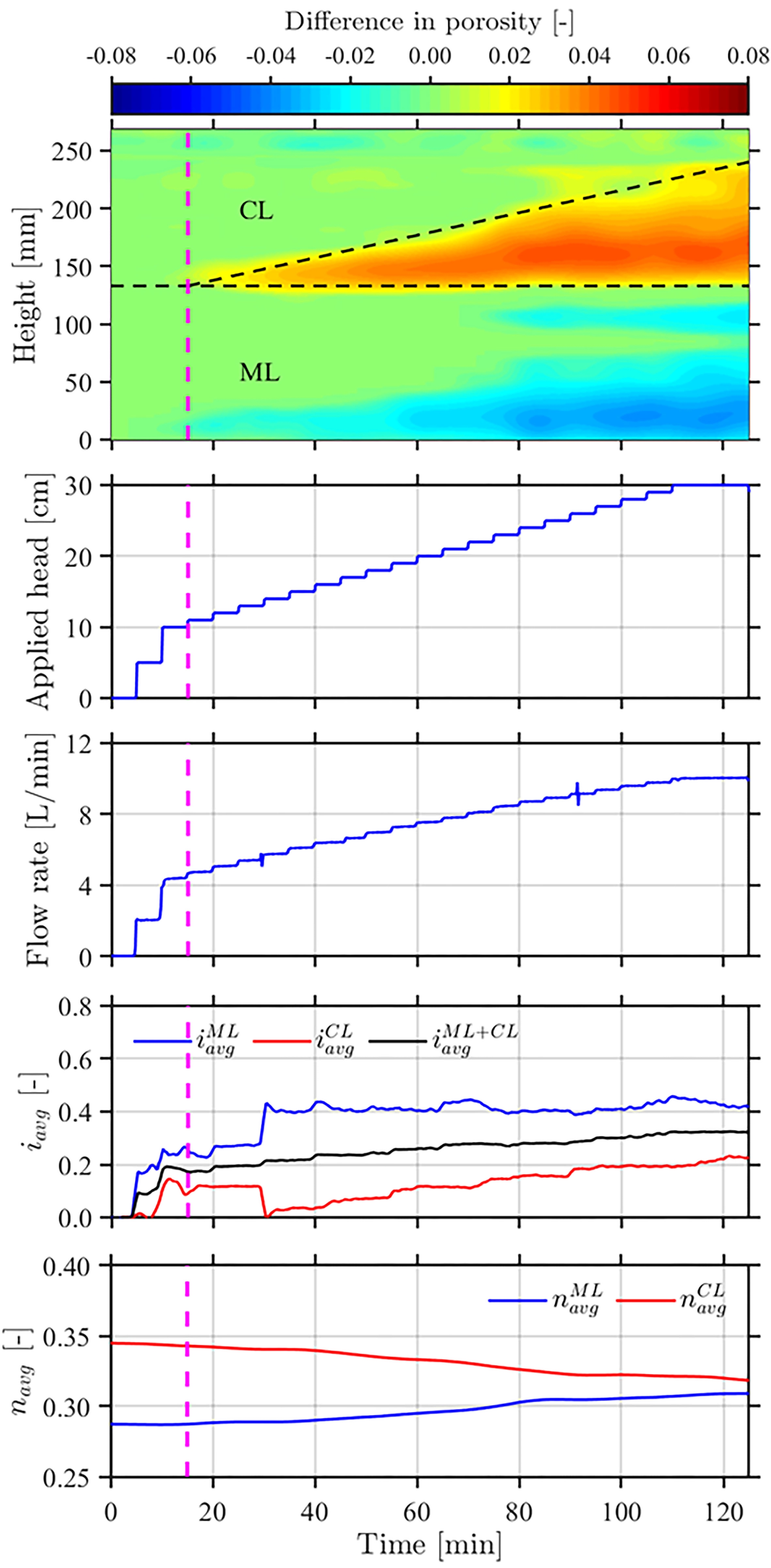}
     \captionof{figure}{Experimental observations for FC15\_LH2 case}
     \label{fig:FC15LH2}
\end{Figure}
\begin{Figure}
     \centering
     \captionsetup{type=figure,justification=centering}
     \includegraphics[width=0.92\textwidth]{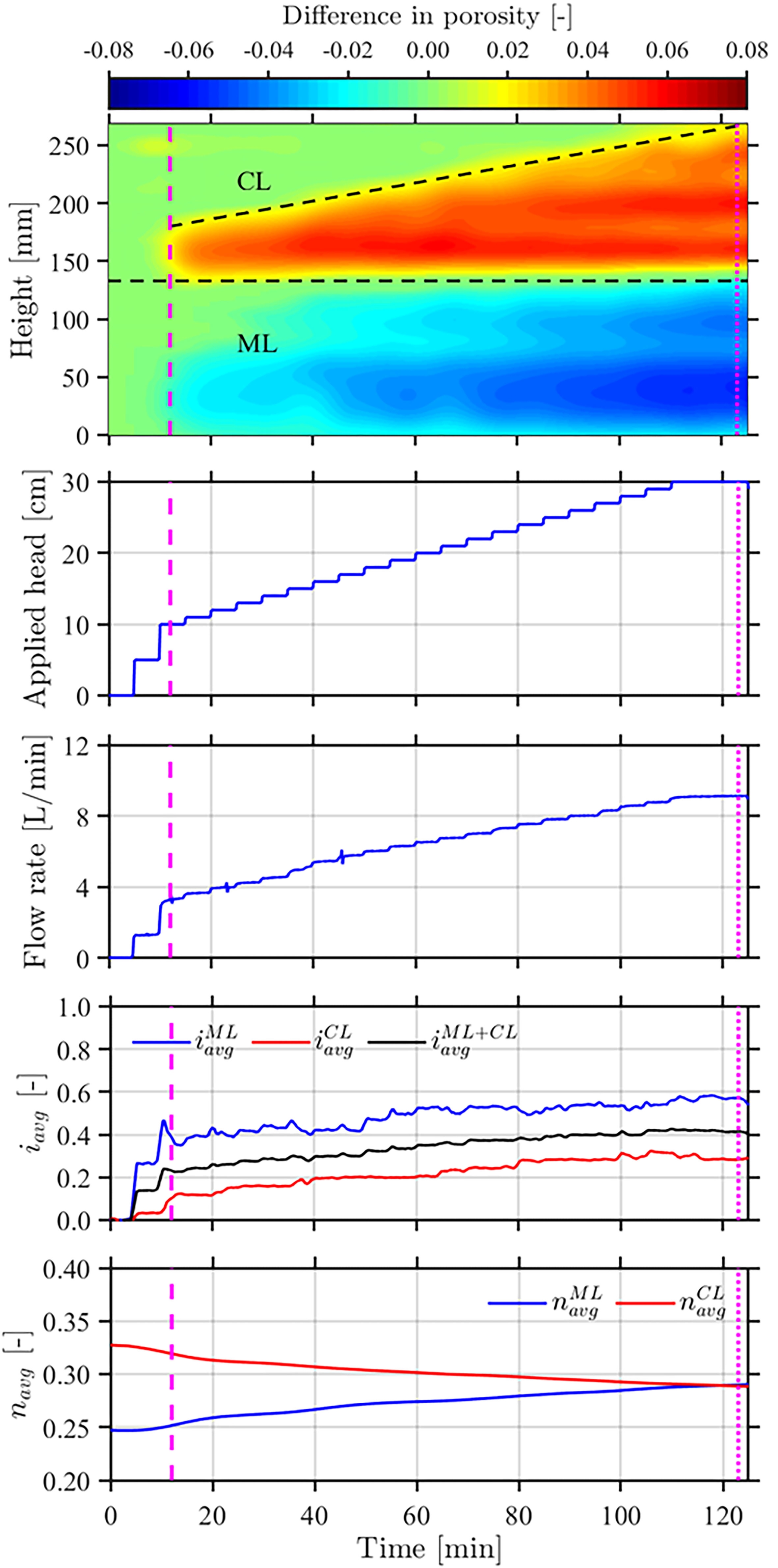}
     \captionof{figure}{Experimental observations for FC20\_LH2 case}
     \label{fig:FC20LH2}
\end{Figure}
\begin{Figure}
     \centering
     \captionsetup{type=figure,justification=centering}
     \includegraphics[width=0.92\textwidth]{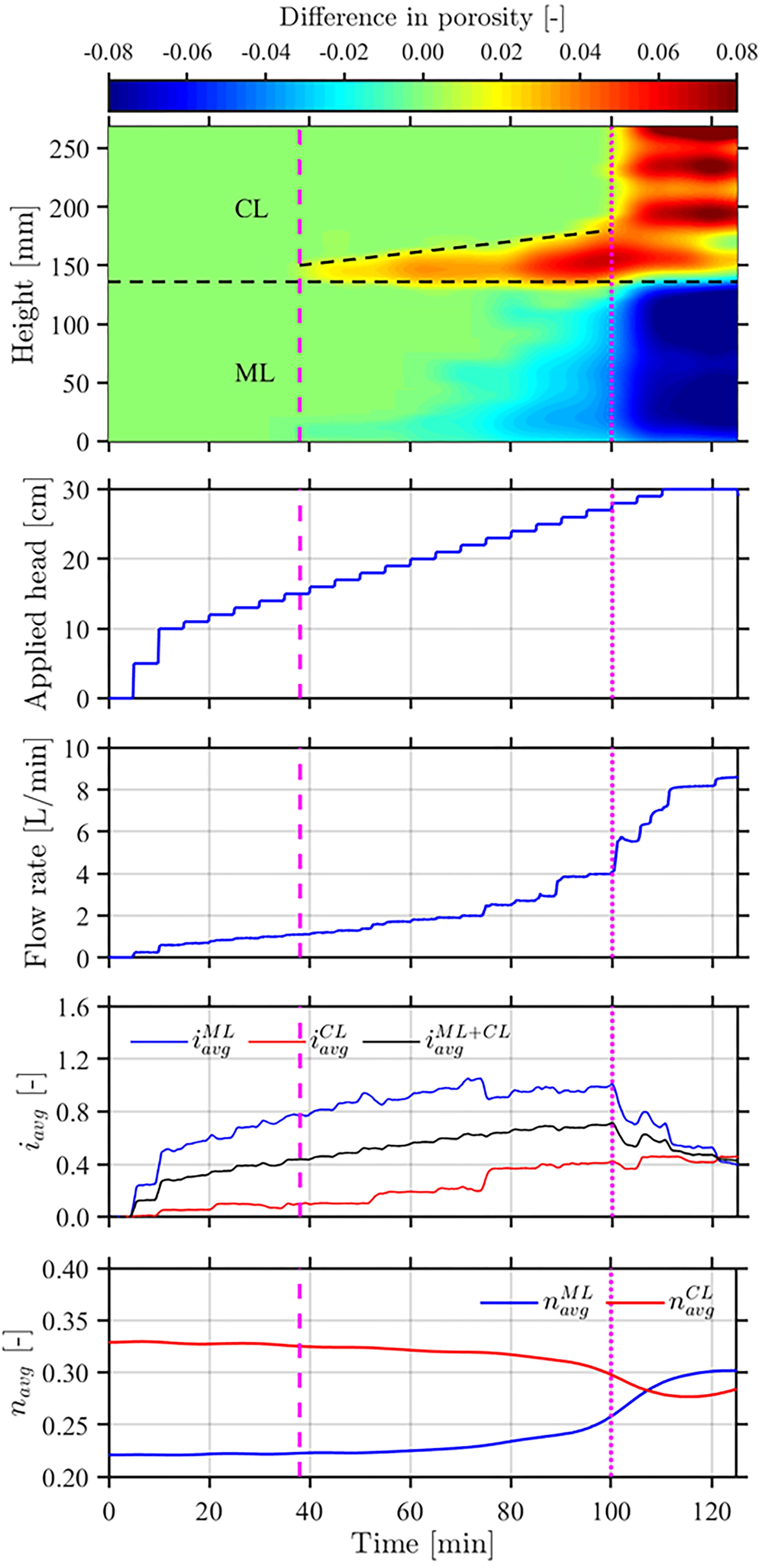}
     \captionof{figure}{Experimental observations for FC25\_LH2 case}
     \label{fig:FC25LH2}
\end{Figure}

\end{multicols}

\end{landscape}

\pagestyle{plain}

\clearpage

\subsection{Based on the conventional approach using flow velocity and average hydraulic gradient}

\begin{figure}[ht]
     \centering
     \includegraphics[width=\textwidth]{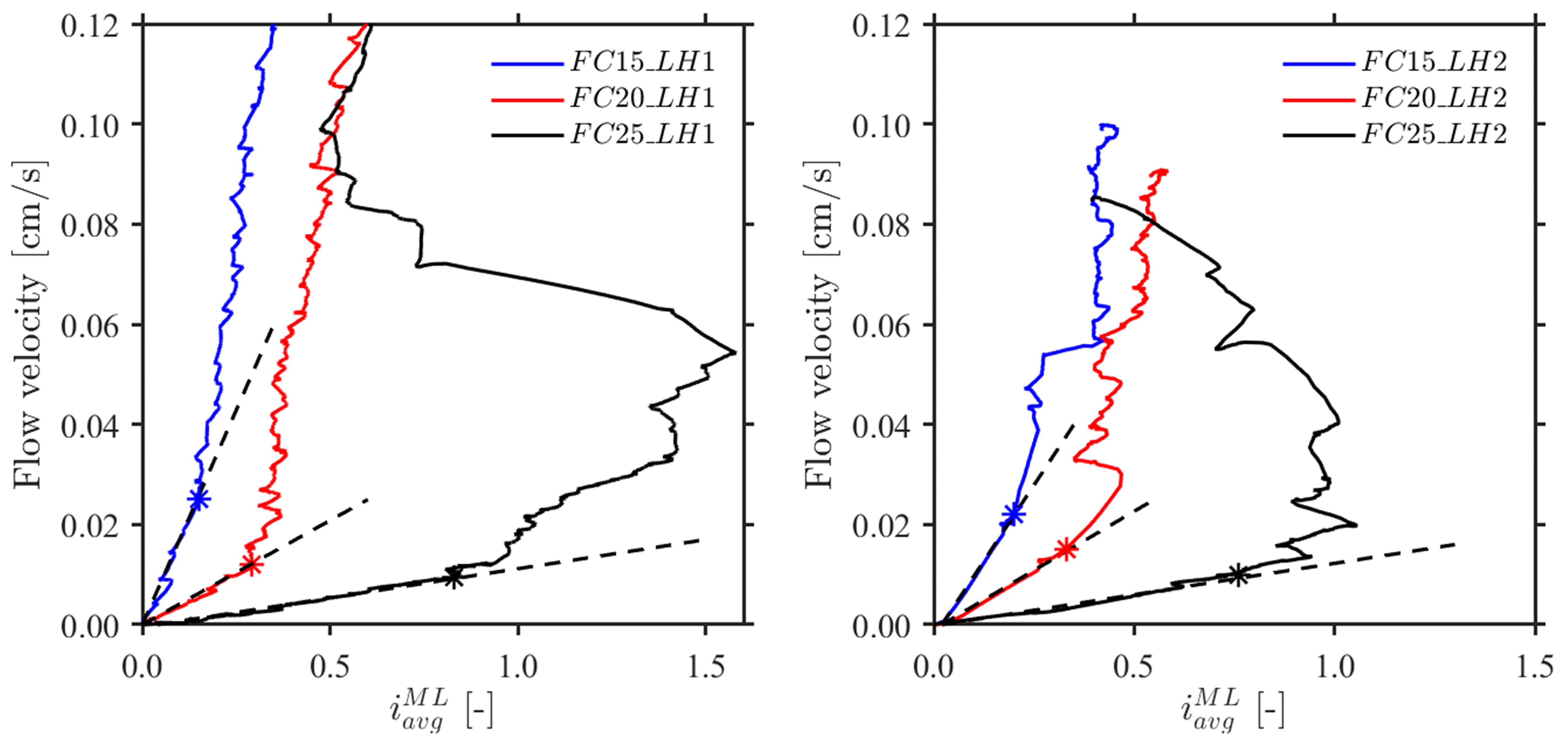}
     \caption{Average hydraulic gradient in the ML against flow velocity for the experiments performed under (a) LH1 and (b) LH2 hydraulic loading conditions}
     \label{fig:i_vs_Vf}
 \end{figure}

The variation of flow velocity with the average hydraulic gradient in the ML is presented in Fig.~\ref{fig:i_vs_Vf}a for LH1 and Fig.~\ref{fig:i_vs_Vf}b for LH2 cases. With increasing applied head, Fig.~\ref{fig:i_vs_Vf} demonstrates that the flow velocity initially increased linearly with increasing $i_{avg}^{ML}$, following Darcy’s law. The onset of suffusion, as indicated with an asterisk mark in Fig.~\ref{fig:i_vs_Vf}, is given by a change in gradient in the flow velocity and average hydraulic gradient curve. Using this conventional approach, the average hydraulic gradient at the onset condition, $i_{onset}^{Darcy}$, is listed in Table~\ref{tab:Table4.4}. Comparable values of $i_{onset}^{TDR}$ and $i_{onset}^{Darcy}$ are noted in Table~\ref{tab:Table4.4}, which validates the applicability of using porosity-based field maps from the spatial TDR approach to identify the limiting onset condition.

According to Darcy’s law, the slope of the flow velocity vs. average hydraulic gradient curves prior to the onset condition is indicative of the average hydraulic conductivity of the mixture layer, $k_{avg}^{Darcy}$, which is listed in Table~\ref{tab:Table4.4} for all cases. Note that the average hydraulic conductivity is unable to account for the spatial variation of local conductivity in the same manner as the modified Kozeny-Carmen expression presented by \citet{annapareddy2022computation}. As expected, the average hydraulic conductivity decreased with increasing fines content, and comparable values were noted for both hydraulic loading conditions. \citet{chapuis2004predicting} presented an empirical expression to estimate the average hydraulic conductivity, $k_{avg}^{Chapuis}$, based on several laboratory experiments, which is given by:
\begin{equation}
    k_{avg}^{Chapuis}=2.46 \left[\frac{d_{10}^2 \left(n_{avg}^{ML}\right)^3}{\left(1-n_{avg}^{ML}\right)^2}\right]^{0.78}
\label{eq:6}
\end{equation}
The average hydraulic conductivity obtained from Eq.~\ref{eq:6} is also listed in Table~\ref{tab:Table4.4} and showed a good agreement with the average hydraulic conductivity obtained in experiments from Fig.~\ref{fig:i_vs_Vf}.

All tested specimens in this study have an underfilled or transitionally underfilled fabric. Therefore, it is expected that $i_{onset}^{Darcy}$ and $i_{onset}^{TDR}$ would be significantly lower than the \citet{terzaghi1939} critical hydraulic gradient for heave failure, $i_{crit}^{Terz}$, which is listed in Table~\ref{tab:Table4.4} and determined based on the average porosity of the mixture layer and a specific gravity of 2.5 for the glass beads. Both $i_{onset}^{Darcy}$ and $i_{onset}^{TDR}$ are significantly lower than $i_{crit}^{Terz}$ and this is attributed to the fact that the stress is predominantly carried by the coarser fraction \citep{skempton1994experiments,shire2014fabric,hunter2018visualisation}. This implies that the effective stress on the finer fraction is only a portion of the effective stress carried by the coarser fraction, which can be described by:
\begin{equation}
    i_{onset}=\alpha \times i_{crit}^{Terz}
\label{eq:7}
\end{equation}
where $i_{onset}$ is the hydraulic gradient at the onset of suffusion observed in the experiments and $\alpha$ is the stress reduction factor which generally ranges between 0 and 1 for suffusive soils, with a larger value of $\alpha$ indicating greater resistance to the onset of suffusion \citep{li2013capillary,hunter2018visualisation}. $\alpha$ values calculated from the spatial TDR approach, $\alpha_{onset}^{TDR}$, and conventional Darcy approach, $\alpha_{onset}^{Darcy}$, are listed in Table~\ref{tab:Table4.5}. $\alpha$ displayed a strong dependence on the fines content, with a larger $\alpha$ observed for higher fines content. An increase in the fines content resulted in $\alpha$ increasing monotonically for the samples considered in this study. This may be attributed to the preparation of the test specimens, which were prepared by premixing the coarser and finer fractions with 4-6\% moisture content and compacted using the moist tamping method. However, further investigations are required to assess the influence of the sample preparation technique on the stress transmission process. Note that a similar trend with $\alpha$ increasing with increasing fines content was observed in the numerical analysis of \citet{shire2014fabric} and \citet{sufian2021influence} which was attributed to a greater proportion of finer particles participating in active stress-carrying roles as fines content increases, leading to greater resistance against the onset of suffusion. Further, $\alpha$ increased by approximately 39\% when the fines content increased from 15\% to 20\%, while it increased by approximately 55\% when the fines content changed from 20\% to 25\%. This signifies a significant increase in $\alpha$ when the soil fabric changes from underfilled to transitionally underfilled. In addition, for specimens with an underfilled fabric ($f_c=15\%$ and 20\%), the $\alpha$ values for LH2 hydraulic loading condition are higher than for LH1, indicating that the critical hydraulic gradient increased as the rate of hydraulic loading increased. However, for specimens with a transitionally underfilled fabric ($f_c=25\%$), both hydraulic loading conditions displayed a similar $\alpha$ value. The higher critical hydraulic gradients at the onset of suffusion with a faster rate of hydraulic loading condition were also noted by \citet{luo2013hydro} and \citet{rochim2017effects}. This can primarily be attributed to higher cumulative energy expended at the onset condition, $E_{onset}$, in LH2 compared to LH1, as listed in Table~\ref{tab:Table4.4}. $E_{onset}$ was calculated by the time integration of total flow power, $P$, from 0 to $t_{onset}$ and the total flow power can be determined as \citep{sibille2015description,marot2016assessing,rochim2017effects}:
 \begin{equation}
     P=Q\gamma_w\Delta h
 \label{eq:8}
 \end{equation}
where $Q$ is the flow rate, $\gamma_w$ is the unit weight of the water and $\Delta h$ is the hydraulic head drop across the upstream and downstream sections of the specimen. However, it is acknowledged that the influence of rate of hydraulic loading on $\alpha$ is based on a limited experimental data set and further investigations are required to better understand the relationship between rate of hydraulic loading and $\alpha$. Existing empirical relationships for $\alpha$ do not suggest any dependence on the rate of hydraulic loading. \citet{li2008seepage} proposed an empirical expression for $\alpha$ as a function of $d_{85}^{FF}$ and $O_{50,CF}^{ML}$:
\begin{equation}
     \alpha_{Li}=3.85 \left(\frac{d_{85}^{FF}}{O_{50,CF}^{ML}}\right)-0.616
\label{eq:9}
\end{equation}
Table~\ref{tab:Table4.5} lists $\alpha_{Li}$ for all cases, noting that $\alpha_{Li}$ is independent of the rate of hydraulic loading. Eq.~\ref{eq:9} was observed to overpredict $\alpha$ at smaller fines content, while it underpredicted $\alpha$ at higher fines content.

\begin{table}[ht]
\caption{Comparison of stress reduction factors calculated in this study and those obtained from Eq.~\ref{eq:9} proposed by \citet{li2008seepage}}
\begin{center}
\begin{tabular}{@{} lcccccc @{}}
\toprule
$\alpha$ & FC15\_LH1 & FC20\_LH1 & FC25\_LH1 & FC15\_LH2 & FC20\_LH2 & FC25\_LH2 \\
\midrule
$\alpha_{Terz}^{TDR}$&0.17&0.27&0.63&0.22&0.33&0.65 \\
$\alpha_{Terz}^{Darcy}$&0.13&0.25&0.67&0.18&0.29&0.62 \\
$\alpha_{Li}$&0.35&0.28&0.28&0.35&0.28&0.28 \\
\bottomrule
\end{tabular}
\end{center}
\label{tab:Table4.5}
\end{table}

\section{Progression of suffusion}

The progression of suffusion is characterised by the increasing infiltration height of the finer fraction into the CL as indicated in the difference in porosity field maps shown in Figs.~\ref{fig:FC15LH1}$-$\ref{fig:FC25LH2}. The rate of infiltration into the CL, $\Dot{H}_{infil}$ (provided in Table~\ref{tab:Table4.4}) can be quantified as the gradient of $H_{infil} (t)$ in the difference in porosity field map. A significantly lower $\Dot{H}_{infil}$ is observed in transitionally underfilled fabrics ($f_c=25\%$) compared to underfilled fabrics ($f_c=15\%$ and 20\%). This is attributed to the differing nature of progression observed in underfilled and transitionally underfilled gap-graded soils. In underfilled soils, the infiltration of the finer fraction gradually leads to the development of a complete mixture condition, while in transitionally underfilled soils, minimal infiltration is observed until a complete mixture zone is formed very rapidly. This can be visualised by comparing the field maps seen in Fig.~\ref{fig:FC20LH1} for FC20\_LH1 and Fig.~\ref{fig:FC25LH1} for FC25\_LH1, as an example.

Figs.~\ref{fig:FC15LH1}$-$\ref{fig:FC25LH2} also show that $n_{avg}^{ML}$ increased with the progression of suffusion, and simultaneously, $n_{avg}^{CL}$ reduced, which is a result of the migration of finer particles from the ML to the CL. The rate of change in $n_{avg}^{ML}$ and $n_{avg}^{CL}$ increased as suffusion progressed, particularly for the specimens with $f_c=25\%$. This indicates that while the infiltration height, $H_{infil} (t)$, increased linearly with the progression of suffusion, the rate of particle migration evolved non-linearly, which is explored further in the subsequent analysis.

The progression of particle migration led to the condition of a complete mixture where the finer fraction infiltrated through the entire CL. This was observed in all cases, except in the case of FC15\_LH2. The time at which the complete mixture layer formed, $t_{CM}$, is indicated with dotted vertical lines in Figs.~\ref{fig:FC15LH1}$-$\ref{fig:FC25LH2}. The time, $t_{CM}$, and the average gradient at the condition of the complete mixture, $i_{CM}$, is listed in Table~\ref{tab:Table4.4}. $i_{CM}$ for $f_{c}=15\%$ and 20\% are moderately greater than $i_{onset}$, while it is significantly higher for $f_{c}=25\%$, which is reflective of the rapid complete mixture condition formed in the specimens with a transitionally underfilled fabric. With LH2, the complete mixture was attained at a much lower hydraulic head for $f_{c}=25\%$, whereas the opposite trend is noted for $f_{c}=15\%$ and 20\%. The flow rate corresponding to the complete mixture condition, $Q_{CM}$, is listed in Table~\ref{tab:Table4.4}, where it is observed that $Q_{CM}$ is significantly higher compared to the flow rate at the onset condition, $Q_{onset}$. The flow rate increased non-linearly between the onset and complete mixing conditions for $f_{c}=25\%$, while a linear increase was noted for $f_{c}=15\%$ and 20\%. The increase in flow rate is primarily governed by the evolution of particle migration between the onset and complete mixing conditions. In the specimens with $f_{c}=25\%$, the particle migration initially evolved slowly from the onset condition, followed by a rapid increase as it approached the complete mixing condition (Figs.~\ref{fig:FC15LH1_mf}a$-$\ref{fig:FC25LH2_mf}a). In contrast, particle migration progressed steadily between onset and complete mixing conditions in the specimens with $f_{c}=15\%$ and 20\% (refer Figs.~\ref{fig:FC15LH1_mf}a$-$\ref{fig:FC20LH1_mf}a and Figs.~\ref{fig:FC15LH2_mf}a$-$\ref{fig:FC20LH2_mf}a). This reflects the difference in the behaviour of underfilled and transitionally underfilled fabrics. This may be attributed to the presence of a greater amount of active finer particles in transitionally underfilled fabric compared to underfilled fabric. The $n_{avg}^{ML}$ and $n_{avg}^{CL}$ approached a similar value at the point of the complete mixture, beyond which the porosity of the ML increased significantly, while the porosity of the CL remained constant or slightly reduced. This is the result of the accumulation of the finer particles on the top of the upper filter once a complete mixture is formed. For $f_{c}=25\%$, a sharp drop in the $i_{avg}^{ML}$ was observed at the complete mixture condition. This is attributed to the rapid migration of finer particles from the ML to the CL. When the finer particles migrated rapidly from the ML, the hydraulic conductivity in this zone increased significantly, which in turn resulted in a reduction in the hydraulic gradient. 

\subsection{Amount and rate of migration of fines}

\begin{figure}[ht]
     \centering
     \includegraphics[width=0.5\textwidth]{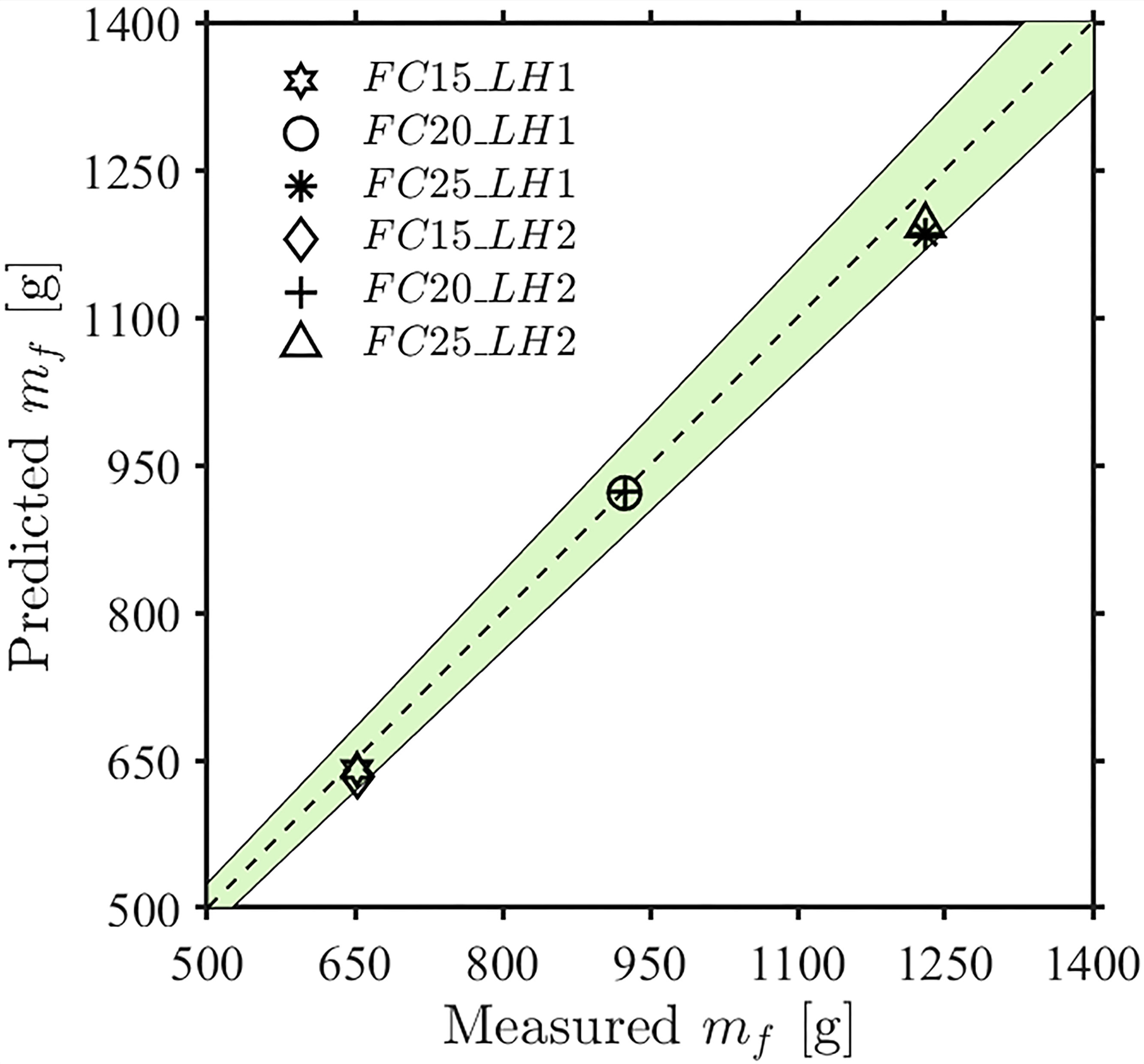}
     \caption{Mass of finer fraction, $m_f$ predicted using Eq.~\ref{eq:10} at time, $t = 0$ is compared against the known $m_f$ measured during specimen preparation stage for all the test cases listed in Table~\ref{tab:Table4.1}}
     \label{fig:Predicted_mf}
 \end{figure}

The temporal variation of $n_{avg}^{ML}$ and $n_{avg}^{CL}$ from spatial TDR can be used to quantify the amount of fines that have migrated from the mixture layer to the coarse layer. The mass of the finer fraction, $m_f$, at a given time can be determined by:
\begin{equation}
    m_f(t)=G_s \gamma_w V_{FF}(t)
\label{eq:10}
\end{equation}
where $G_s$ is the specific gravity of the glass beads, $\gamma_w$ is the unit weight of water and $V_{FF}(t)$ is the solid volume of the finer fraction, which can be obtained using: 
\begin{equation}
    V_{FF}(t)=\left[1-n_{avg,FF}(t)\right]\times V
\label{eq:11}
\end{equation}
where $V$ is the total volume and $n_{avg,FF}(t)$ is the average porosity of the finer fraction at a given time, which can be obtained using the following expression given by \citet{annapareddy2022computation}:
\begin{equation}
   n_{avg,FF}(t)=n_{avg}(t)-n_{avg,CF}(t)+1 
\label{eq:12}
\end{equation}
where $n_{avg,CF}(t)$ is the average porosity of the coarser fraction at a given time, $t$, and $n_{avg}(t)$ is the average porosity obtained from spatial TDR data. $n_{avg,CF}(t)$ is given by \citep{Kovacs1981}:
\begin{equation}
    n_{avg,CF}(t)=n_{avg}(t)+\frac{f_c}{100}\left[1-n_{avg}(t)\right]
\label{eq:13}
\end{equation}
The residual mass of the finer fraction in the ML, $m_f^{ML}$, at a given time can be obtained by using the average porosity of the ML, that is, $n_{avg}=n_{avg}^{ML}$ in Eq.~\ref{eq:12} and \ref{eq:13}. Similarly, using the average porosity of the CL (i.e., $n_{avg}=n_{avg}^{CL}$), provides the mass of finer fraction migrated into the CL, $m_f^{CL}$. The initial mass of the finer fraction in the ML predicted from Eq.~\ref{eq:10} is compared against the known mass of finer friction that was initially added to the ML in each experiment in Fig.~\ref{fig:Predicted_mf}. The dashed line in Fig.~\ref{fig:Predicted_mf} represents a 1:1 line, while the solid lines denote a $\pm10\%$ error band. Eq.~\ref{eq:10} slightly underestimated the initial mass of the finer fraction, which was attributed to measurement errors associated with spatial TDR and loss of fines during sample preparation. The results from Eq.~\ref{eq:10} were minimally influenced by the potential presence of preferential flow paths as these mainly relied on the measurements of local porosity from spatial TDR.

\begin{figure}[ht]
     \centering
     \includegraphics[width=0.55\textwidth]{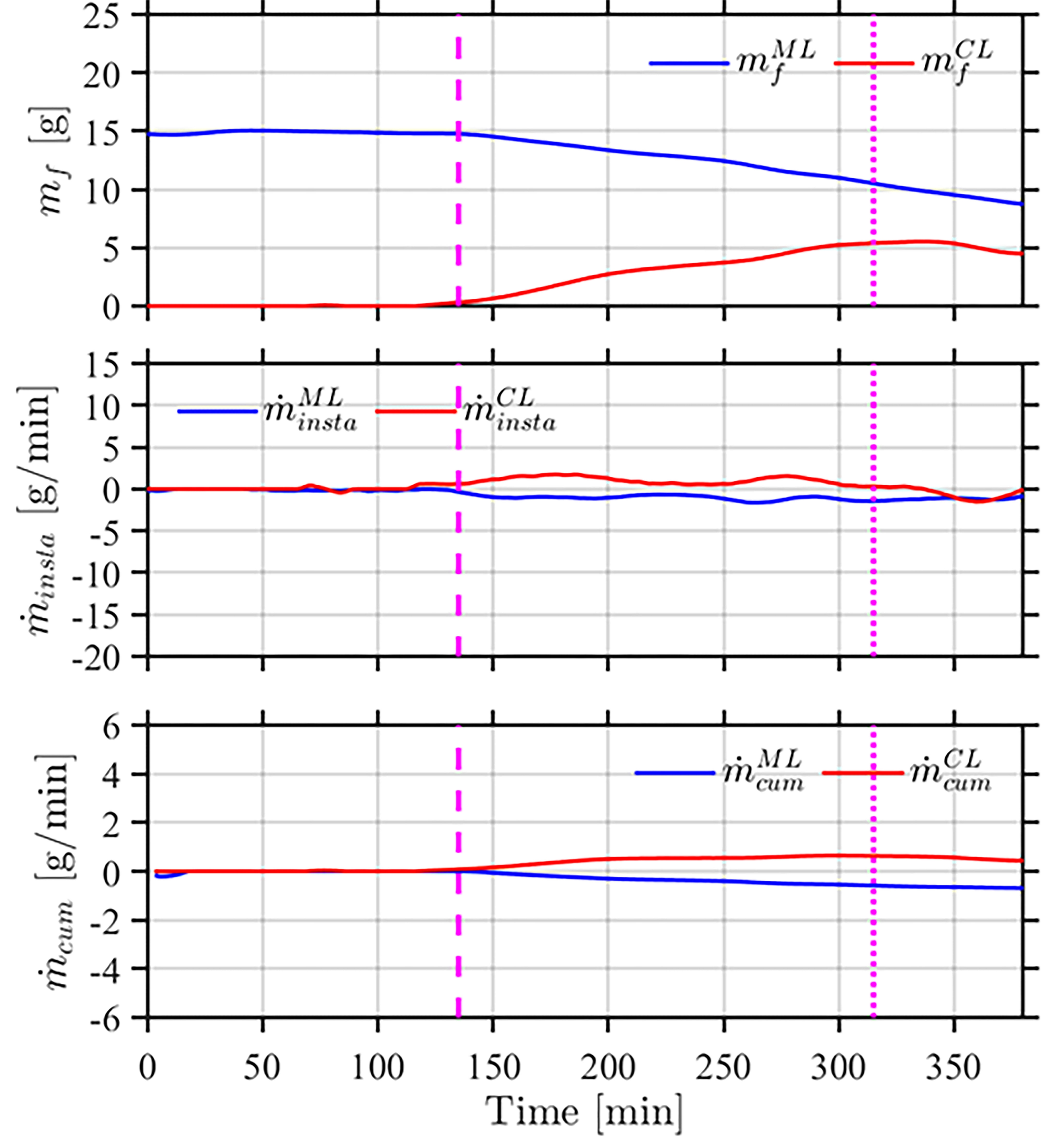}
     \caption{Changes in the (a) mass of finer fraction, $m_f$ (b) instantaneous rate of change in the mf and (c) cumulative rate of change in the $m_f$ for FC15\_LH1. In Figs.~\ref{fig:FC15LH1_mf}$-$\ref{fig:FC25LH2_mf}, the limiting onset condition is indicated by dashed line and the complete mixture condition is indicated by dotted line.}
     \label{fig:FC15LH1_mf}
 \end{figure}

\begin{figure}[ht]
     \centering
     \includegraphics[width=0.55\textwidth]{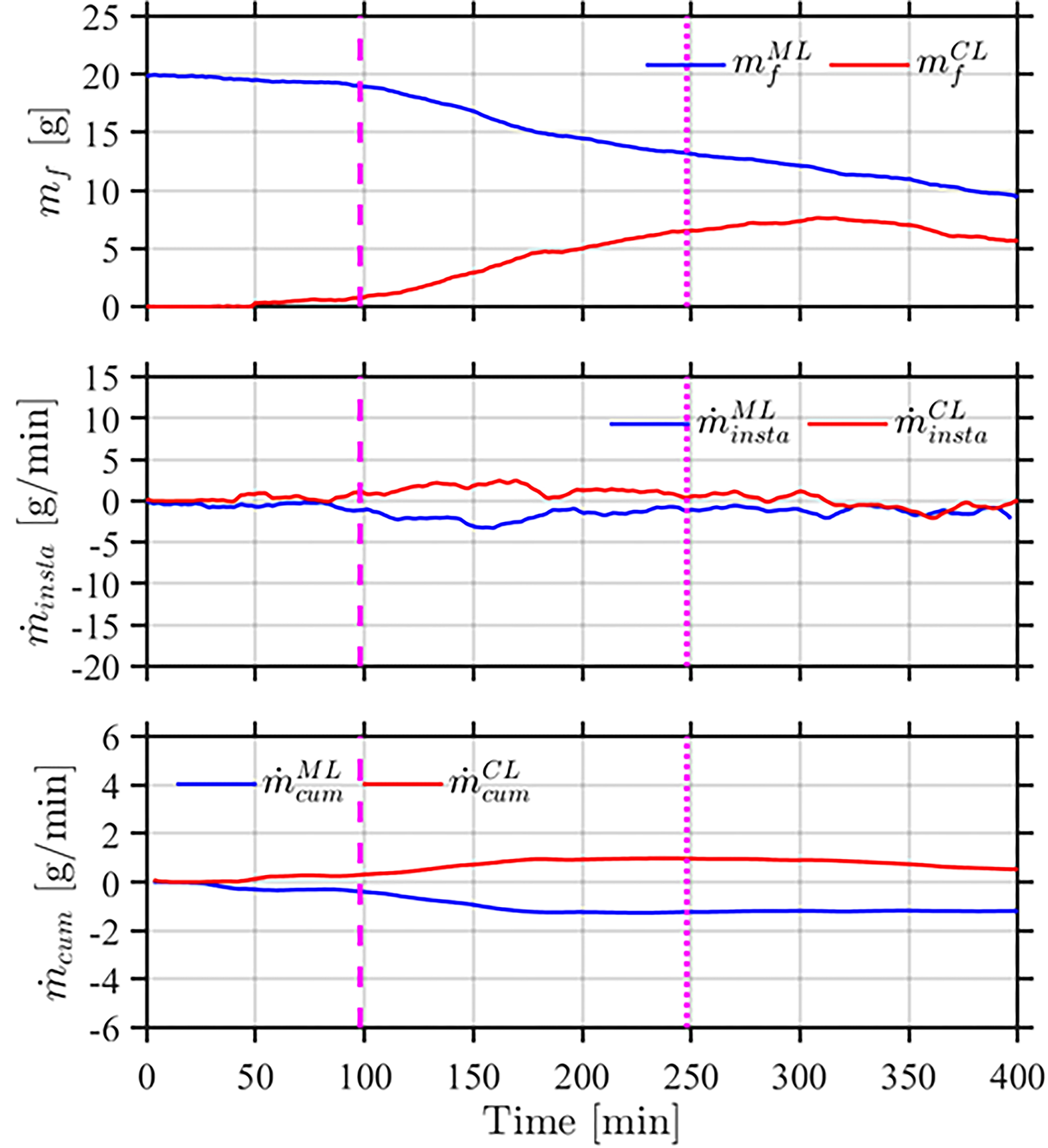}
     \caption{Changes in the (a) mass of finer fraction, $m_f$ (b) instantaneous rate of change in the $m_f$ and (c) cumulative rate of change in the $m_f$ for FC20\_LH1.}
     \label{fig:FC20LH1_mf}
 \end{figure}

\begin{figure}[ht]
     \centering
     \includegraphics[width=0.55\textwidth]{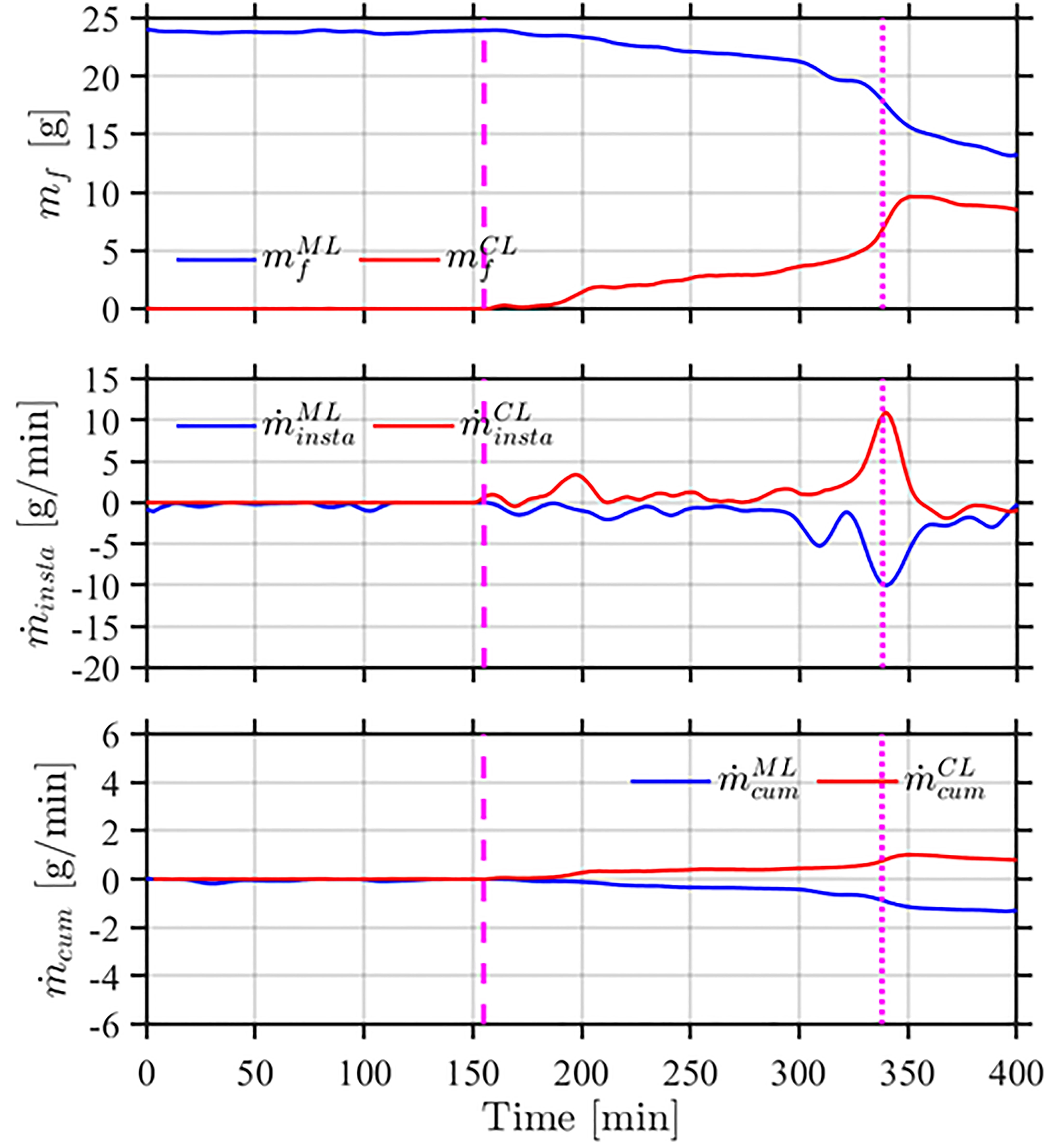}
     \caption{Changes in the (a) mass of finer fraction, $m_f$ (b) instantaneous rate of change in the $m_f$ and (c) cumulative rate of change in the $m_f$ for FC25\_LH1.}
     \label{fig:FC25LH1_mf}
 \end{figure}

\begin{figure}[ht]
     \centering
     \includegraphics[width=0.55\textwidth]{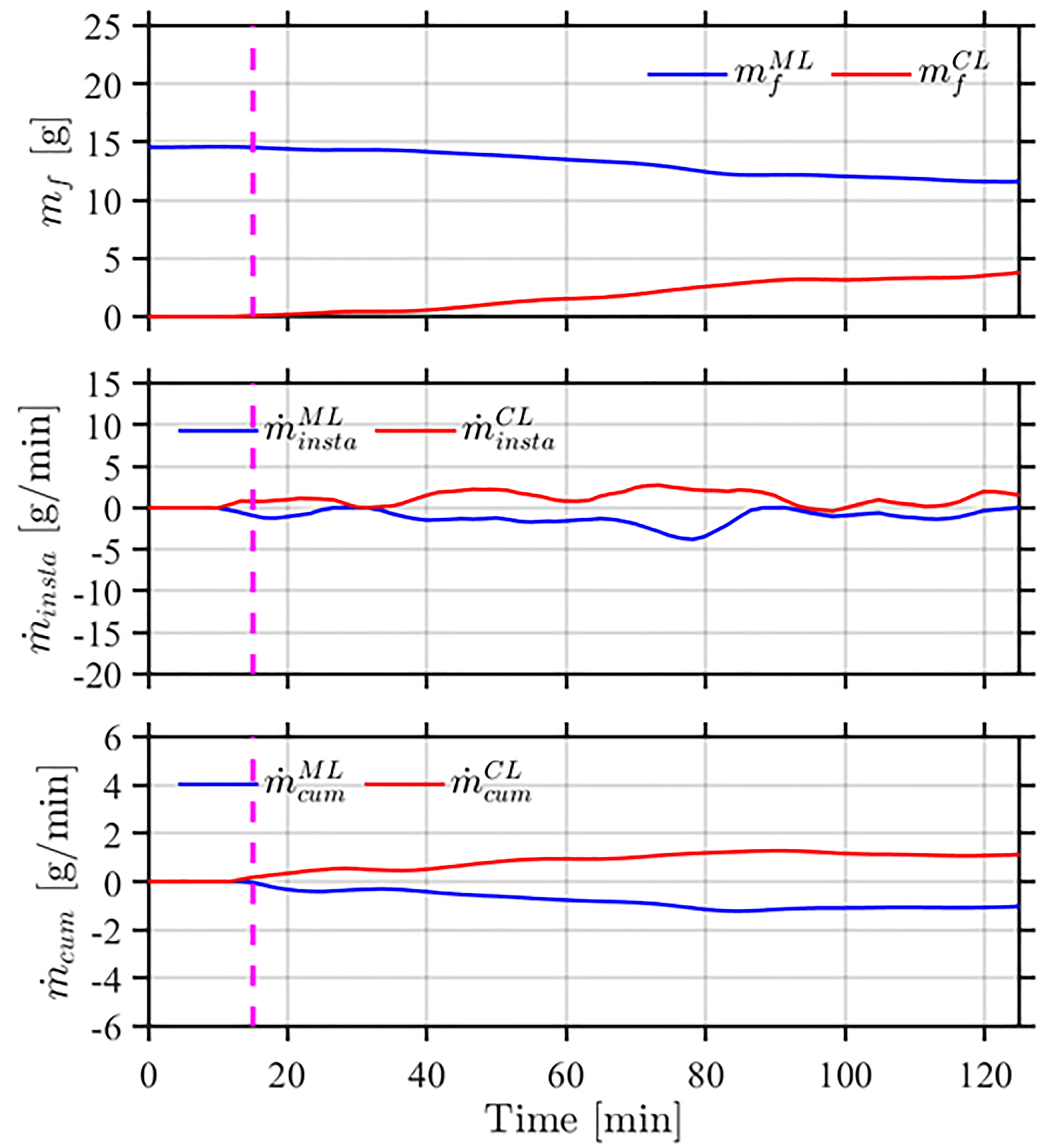}
     \caption{Changes in the (a) mass of finer fraction, $m_f$ (b) instantaneous rate of change in the $m_f$ and (c) cumulative rate of change in the $m_f$ for FC15\_LH2.}
     \label{fig:FC15LH2_mf}
 \end{figure}

\begin{figure}[ht]
     \centering
     \includegraphics[width=0.55\textwidth]{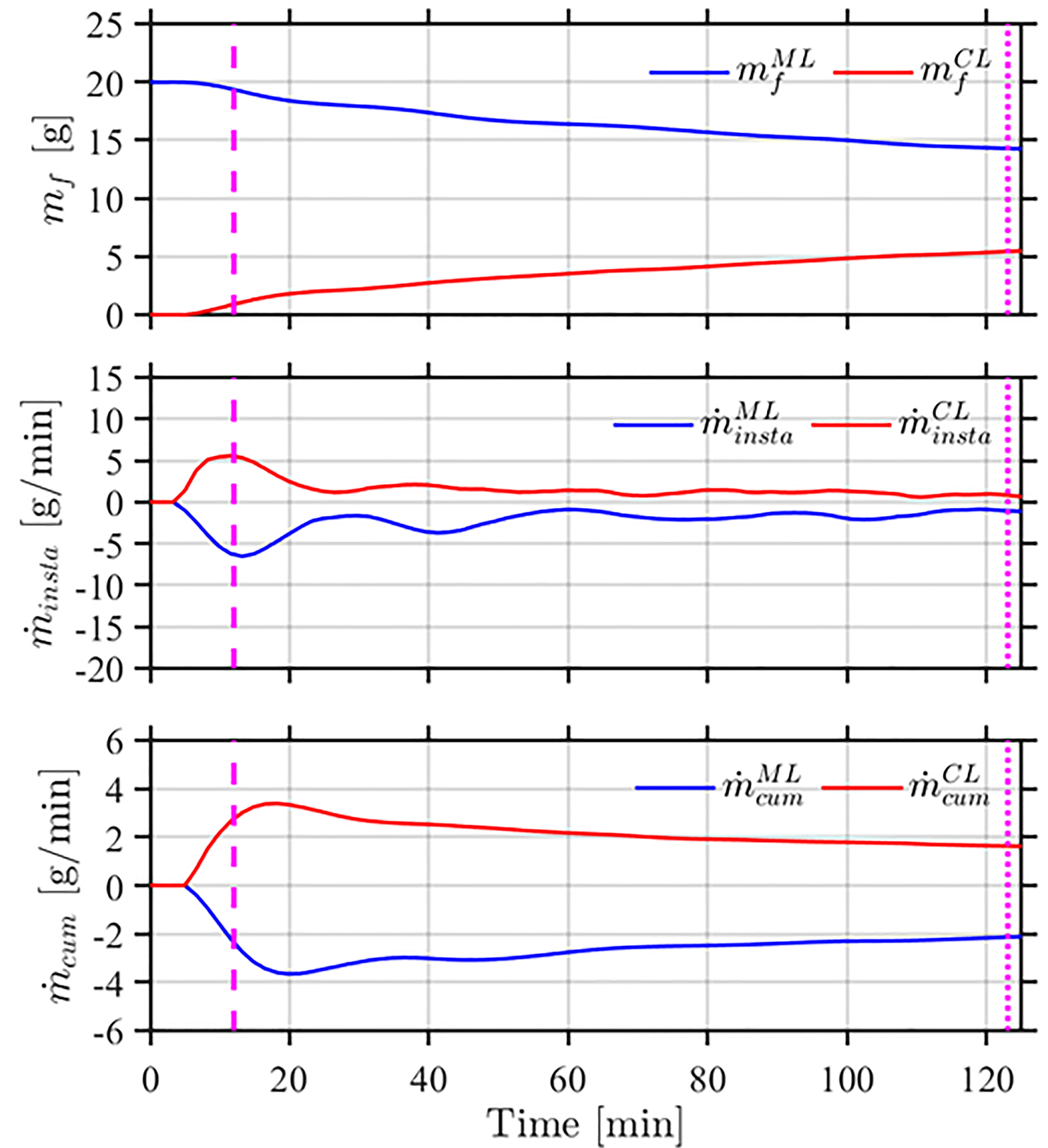}
     \caption{Changes in the (a) mass of finer fraction, $m_f$ (b) instantaneous rate of change in the $m_f$ and (c) cumulative rate of change in the $m_f$ for FC20\_LH2.}
     \label{fig:FC20LH2_mf}
 \end{figure}

\begin{figure}[ht]
     \centering
     \includegraphics[width=0.55\textwidth]{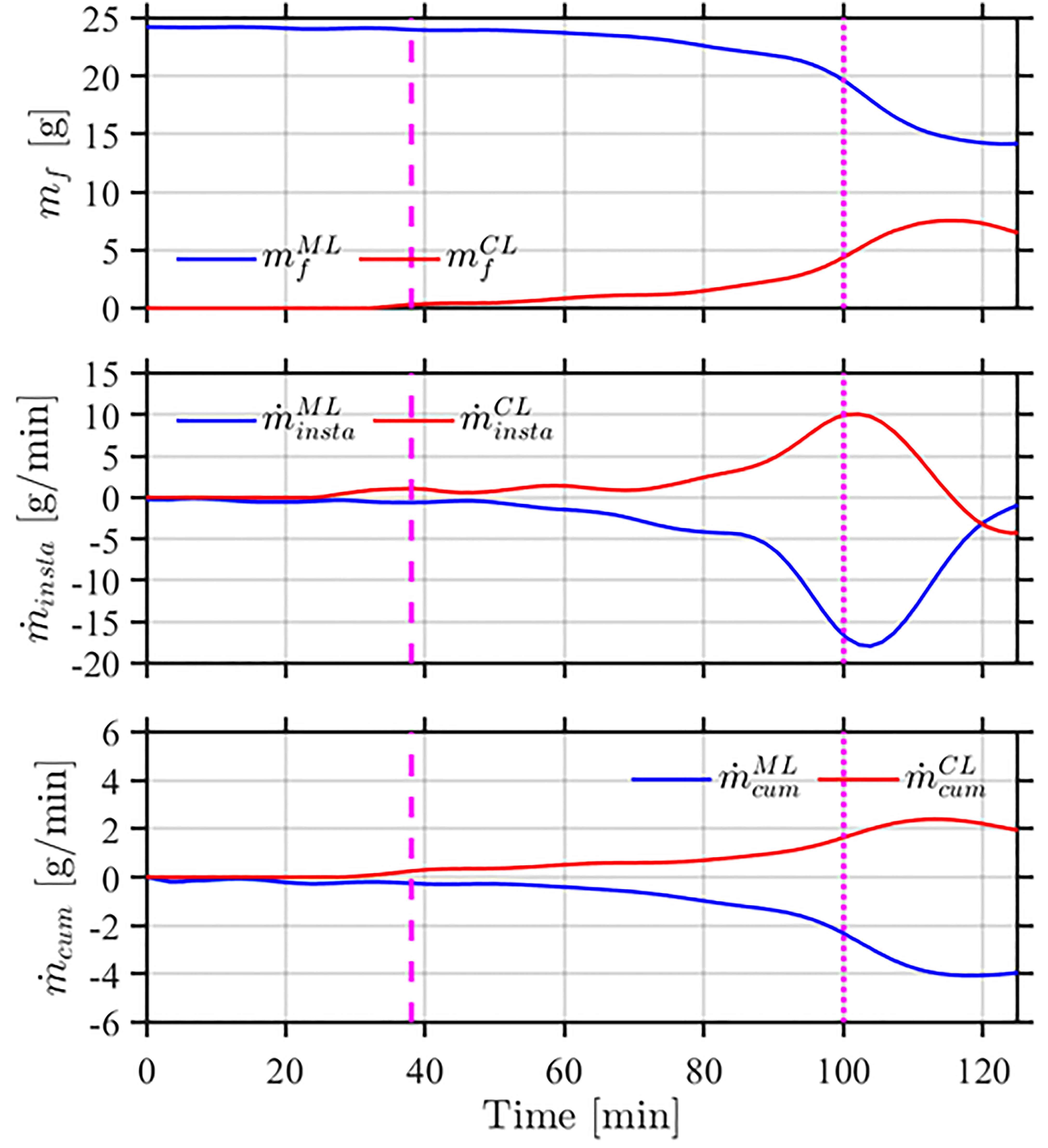}
     \caption{Changes in the (a) mass of finer fraction, $m_f$ (b) instantaneous rate of change in the $m_f$ and (c) cumulative rate of change in the $m_f$ for FC25\_LH2.}
     \label{fig:FC25LH2_mf}
 \end{figure}

The evolution of the mass of the finer fraction in the ML and CL is shown in Figs.~\ref{fig:FC15LH1_mf}a$-$\ref{fig:FC25LH2_mf}a for all cases. The dashed vertical line in Figs.~\ref{fig:FC15LH1_mf}--\ref{fig:FC25LH2_mf}, indicates $t_{onset}$ and is based on the spatial TDR approach, while the dotted vertical line indicates $t_{CM}$. When $t<t_{onset}$, $m_f^{ML}$ is approximately constant equalling to initial fines content present in each respective test listed in Table~\ref{tab:Table4.1} and $m_f^{CL}$ is approximately zero as no particle migration has occurred. For $t_{onset}<t<t_{CM}$, a significant loss of finer fraction is noted in the ML with a subsequent gain of finer fraction into the CL. For $t>t_{CM}$, the loss of finer fraction from ML continues, while the mass of the finer fraction in the CL plateaus. This is because the finer fraction has migrated to the top of the CL at $t=t_{CM}$, and for $t>t_{CM}$, the finer fraction leaves the CL into the upper filter.

The instantaneous rate of change in the mass of the finer fraction in the ML ($\Dot{m}_{insta}^{ML}$) and the CL ($\Dot{m}_{insta}^{CL}$) can be obtained from the gradients of $m_{f}^{ML}$ and $m_{f}^{CL}$. The mass of the finer fraction transported from ML at any time, $t$, can be obtained by deducting the initial mass $m_{f}^{ML}(t=0)$ from $m_{f}^{ML}(t)$. This provides the mass loss of finer fraction from ML. The cumulative loss rate of the finer fraction from ML, $\Dot{m}_{cum}^{ML}$, and the cumulative gain rate of the finer fraction into CL, $\Dot{m}_{cum}^{CL}$, are obtained by considering the cumulative rate of change of the finer fraction in their respective layers. Figs.~\ref{fig:FC15LH1_mf}b$-$\ref{fig:FC25LH2_mf}b show the instantaneous rate of migration of the finer fraction, while the cumulative rate of migration of the finer fraction is displayed in Figs.~\ref{fig:FC15LH1_mf}c$-$\ref{fig:FC25LH2_mf}c for all cases. Positive values of $\Dot{m}_{insta}$ or $\Dot{m}_{cum}$ in Figs.~\ref{fig:FC15LH1_mf}$-$\ref{fig:FC25LH2_mf} indicate a gain in the finer fraction, while the negative values indicate a loss in the finer fraction.

When $t<t_{onset}$, all different rates of migration ($\Dot{m}_{insta}^{ML}$, $\Dot{m}_{insta}^{CL}$, $\Dot{m}_{cum}^{ML}$ and $\Dot{m}_{cum}^{CL}$) are approximately zero it confirms that no migration of the finer fraction has occurred during this stage. The maximum $\Dot{m}_{insta}^{CL}$ and minimum $\Dot{m}_{insta}^{ML}$ were attained in between $t_{onset}$ and $t_{CM}$ except for the cases with $f_c=25\%$, where the maximum $\Dot{m}_{insta}^{CL}$ and minimum $\Dot{m}_{insta}^{ML}$ were attained just after $t_{CM}$ as can be seen in Figs.~\ref{fig:FC25LH1_mf} and \ref{fig:FC25LH2_mf}. $\Dot{m}_{cum}^{CL}$ increased between $t_{onset}<t<t_{CM}$, while simultaneously, $\Dot{m}_{cum}^{ML}$ decreased, indicating that the rate of particle migration was not constant during the progression of suffusion, but rather increased significantly as it approached the complete mixing condition. When $t>t_{CM}$, $\Dot{m}_{cum}^{ML}$ was relatively constant, while a slight reduction was observed in $\Dot{m}_{cum}^{ML}$.

The driving hydraulic gradient at the complete mixture condition, $\Delta \left(i_{avg}^{ML} - i_{onset}^{ML}\right)_{CM}$, average power, $\Bar{P}$ was calculated by averaging the total flow power obtained from Eq.~\ref{eq:8}, amount of loss of finer fraction, $\delta m_f^{ML}$, maximum instantaneous, $\Dot{m}_{insta,max}^{ML}$ and cumulative, $\Dot{m}_{cum,max}^{ML}$ rates of particle migration in the ML between the onset and complete mixing conditions are given in Table~\ref{tab:Table4.6} for all the cases listed in Table~\ref{tab:Table4.1}. Note that in FC15\_LH2, the complete mixture condition was not attained under the applied LH2 hydraulic loading condition. While the amount of loss of finer fraction in LH2 is lower than in LH1, the instantaneous and cumulative rates of migration are significantly higher (approximately 2 to 3 times) in LH2. This is because the average power applied during the progression phase in LH2 is significantly greater than in LH1. Similar observations were noted in \citet{rochim2017effects}. For LH1, the driving hydraulic gradient required in underfilled fabric ($f_c=15\%$ and 20\%) is considerably lower than in transitionally underfilled fabric ($f_c=25\%$) to cause similar amounts of loss of finer fraction. This implies that once the onset condition is attained, a small increment in the local hydraulic gradient can cause a significant amount of particle migration for soils with an underfilled fabric. On the other hand, a substantial increase in the local hydraulic gradient is required to cause particle migration for soils with a transitionally underfilled fabric. This is attributed to the low stress experienced by the finer fraction for soils with an underfilled fabric compared to that in soils with a transitionally underfilled fabric. However, a differing observation is noted for LH2. This may be a result of the large increments in the applied hydraulic gradient at the beginning of LH2 cases, which affected the determination of $i_{onset}^{ML}$ for FC20\_LH2. \par

\begin{table}[ht]
\caption{Key characteristics of particle migration during the progression of suffusion}
\begin{center}
\begin{tabular}{@{} lp{3cm}p{1cm}p{1cm}p{1.7cm}p{1.7cm} @{}}
\toprule
Test ID & $\Delta \left(i_{avg}^{ML}-i_{onset}^{ML}\right)_{CM}$ [--]& $\Bar{P}$ [J/min]& $\Delta m_f^{ML}$ [\%]& $\Dot{m}_{insta, max}^{ML}$ [g/min]& $\Dot{m}_{cum, max}^{ML}$ [g/min] \\
\midrule
FC15\_LH1&0.09&0.40&4.26&1.63&0.58 \\
FC20\_LH1&0.14&0.30&5.79&3.27&1.26 \\
FC25\_LH1&0.79&0.49&6.15&9.97&0.88 \\
FC15\_LH2&–-&–-&–-&-–&-– \\
FC20\_LH2&0.20&1.95&5.15&6.51&3.65 \\
FC25\_LH2&0.26&1.39&3.41&13.87&1.93 \\
\bottomrule
\end{tabular}
\end{center}
\label{tab:Table4.6}
\end{table}

\clearpage 

\begin{figure}[ht]
     \centering
     \includegraphics[width=0.6\textwidth]{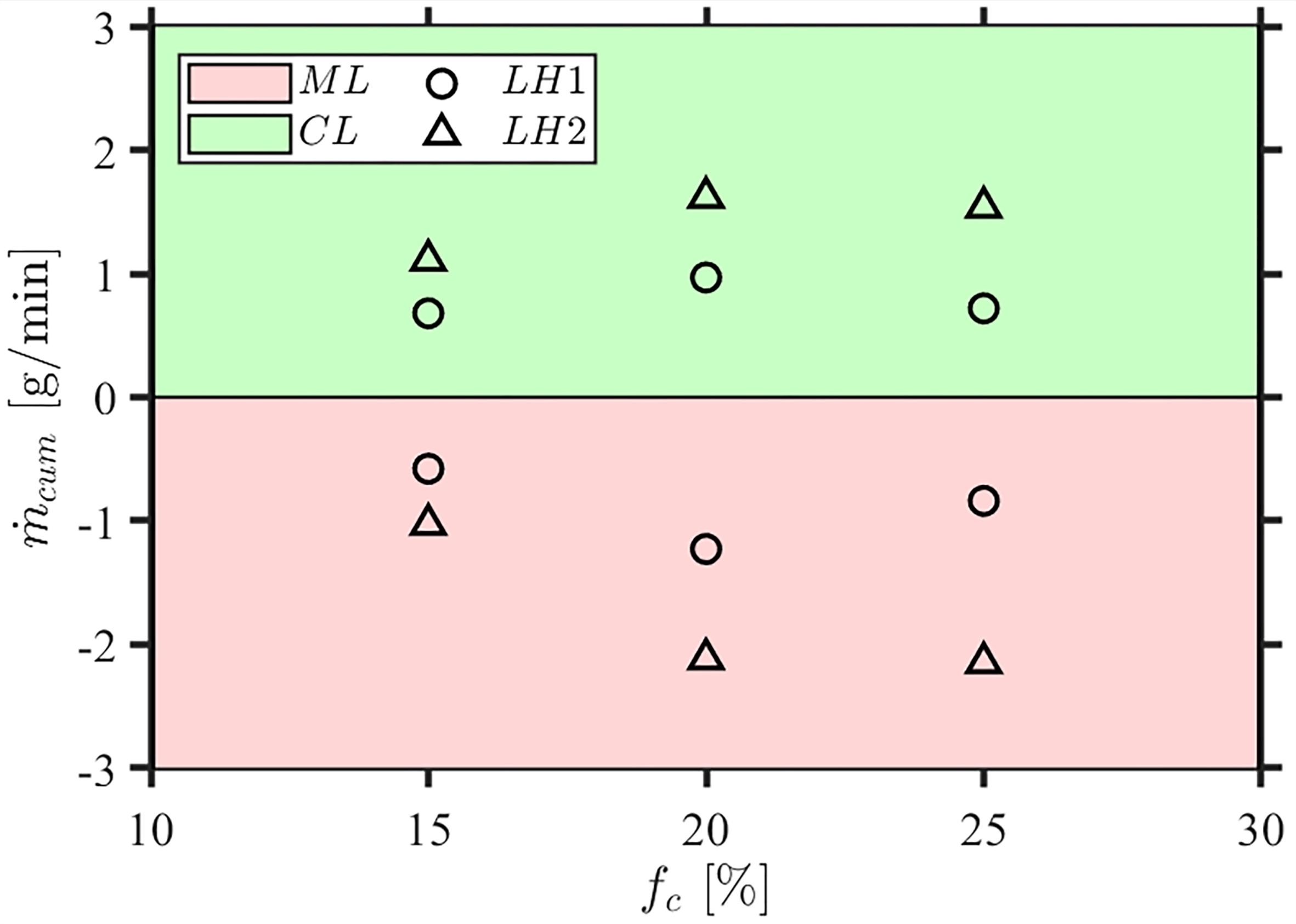}
     \caption{Influence of fines content and rate of hydraulic loading on the cumulative rate of change in the mass of finer fraction at complete mixture condition.}
     \label{fig:ResultSummary}
 \end{figure}

The influence of the fines content and the rate of loading on the cumulative rate of fines migration corresponding to $t_{CM}$ (at the condition of complete mixture) is shown in Fig.~\ref{fig:ResultSummary}. It is observed that $\Dot{m}_{cum}^{ML}$ and $\Dot{m}_{cum}^{CL}$ increased by approximately 52\% and 30\%, respectively when the fines content is increased from 15\% to 20\% for both LH1 and LH2. For LH1, a further increase in the fines content to 25\% resulted in a significant decrease in $\Dot{m}_{cum}^{ML}$ and $\Dot{m}_{cum}^{CL}$ by approximately 46\% and 34\%, respectively, while the $\Dot{m}_{cum}^{ML}$ and $\Dot{m}_{cum}^{CL}$ only reduced slightly for LH2. This is because the maximum and minimum rate of migration at a given time occurred after the attainment of the complete mixture condition for $f_c=25\%$ (refer to Figs.~\ref{fig:FC25LH1_mf}b and \ref{fig:FC25LH2_mf}b), while it is occurred in between the onset and complete mixture conditions for $f_c=15\%$ and 20\% (refer Figs.~\ref{fig:FC15LH1_mf}b$-$\ref{fig:FC20LH1_mf}b and \ref{fig:FC15LH2_mf}b$-$\ref{fig:FC20LH2_mf}b). For the specimens with $f_c=25\%$, the maximum and minimum rate of migration in LH2 occurred further away from the complete mixing condition compared to that in LH1. The faster rate of hydraulic loading yielded higher values of $\Dot{m}_{cum}^{ML}$ and $\Dot{m}_{cum}^{CL}$. The $\Dot{m}_{cum}^{ML}$ and $\Dot{m}_{cum}^{CL}$ in LH2 are approximately 40\% higher than in LH1 for $f_c=15\%$ and 20\%, while they are approximately 57\% greater for $f_c=25\%$. This observation is consistent with \citet{rochim2017effects} and the reason for this has been detailed above.

\section{Conclusions}

A purpose-built coaxial permeameter was employed to explore the spatial and temporal evolution of particle migration during the complete process of suffusion from onset to progression to washout. The influence of fines content in gap-graded granular soils and the rate of hydraulic loading on particle migration was investigated for an upward seepage flow inside a coaxial permeameter cell. The cell acted as a transmission line facilitating electromagnetic measurements using spatial time-domain reflectometry from which the spatial and temporal evolution of local porosity along the height of the specimen were obtained using an inversion technique. In addition, the flow rate and hydraulic gradients within the specimen were monitored using a flow meter and pressure transducers, respectively. The test specimen consists of a mixture layer overlain by a coarse layer of similar height. \par 

A field map of the difference in porosity was generated by considering the changes in the local porosity profile with the initial porosity profile. The limiting onset condition for suffusion was inferred from the difference in porosity field map, from which the critical hydraulic gradient was obtained. The progression of suffusion was characterised by the infiltration of the finer fraction into the coarse layer with the continuation of particle migration leading to the formation of a complete mixture condition. The amount and rate of particle migration within the specimen were determined from the measured porosity profile. The rate of loss of finer particles from the mixture layer and the rate of gain of finer particles into the coarse layer at any given time was presented for each experiment using instantaneous and cumulative rates of particle migration. The key findings from this study are:
\begin{enumerate}
    \item The limiting onset condition for suffusion was readily obtained from the difference in porosity field maps obtained from spatial TDR data. The corresponding critical hydraulic gradients exhibited a close agreement with the critical hydraulic gradients obtained from the conventional approach using average hydraulic gradient and flow velocity curves. This provided strong validation for the applicability of the field maps obtained from spatial TDR for identifying the onset of suffusion irrespective of the filling degree.
    \item The critical hydraulic gradient and critical flow rate showed strong dependence on the fines content. An increase in the fines content yielded higher critical gradients and lower critical flow rates. In contrast, the rate of hydraulic loading showed minimal to negligible influence on the critical hydraulic gradient and critical flow rate, with slightly higher critical hydraulic gradients for the faster rate of loading in underfilled fabrics.
    \item The rate of infiltration of the finer fraction into the coarse layer is significantly lower in transitionally underfilled fabrics ($f_c = 25\%$) compared to underfilled fabrics ($f_c = 15\%$ and 20\%). This is attributed to the differing nature of progression observed in underfilled and transitionally underfilled gap-graded soils. In underfilled soils, the mixture zone gradually developed into the complete mixture condition, while in transitionally underfilled soils, the mixture zone only minimally developed until a complete mixture zone formed very rapidly. The mechanisms preventing the continuation of fines migration and causing the rapid development of a mixture zone require further investigation. However, it can be hypothesised that the controlling factor for the occurrence of this observation is the concentration of fines in the mixture zone reaching a transitional concentration where the fines start to contribute to the transfer of the effective stress.
    \item While the infiltration of the finer fraction increased linearly with the progression of suffusion, the rate of particle migration evolved non-linearly. This indicates the heterogeneous nature of particle migration along the mixture layer, which is the subject of ongoing research.
    \item he rate of particle migration displayed a strong dependence on fines content and rate of hydraulic loading. It was observed that the increase in the rate of particle migration with the increase in fines content reached a plateau when the soil fabric changed from underfilled to transitionally underfilled. This is because in a transitionally underfilled fabric significant portion of the finer fraction participates in the transmission of effective stresses. Further, the faster rate of hydraulic loading resulted in a higher rate of particle migration. However, the preliminary results presented on the influence of the rate of hydraulic loading are based on a limited amount of data and further investigations at differing hydraulic loading conditions are required to better understand the influence of the rate of hydraulic loading on the suffusion process.
\end{enumerate}

These observations on the spatial and temporal evolution of local porosity and particle migration demonstrate the capability of the spatial TDR approach to investigate the entire process of suffusion from onset to progression to washout. Ongoing research aims to utilise these observations on particle migration to evaluate the influence of heterogeneity on particle migration during suffusion processes. While this study was limited to considering idealised gap-graded soils with an underfilled or transitionally underfilled fabric, future research aims to extend the coaxial permeameter cell to systematically investigate real soils considering a wider range of PSDs including broadly graded soils, different particle shapes, hydraulic loading conditions and effective stress conditions.

\section*{Acknowledgements}
This work was funded by an Australian Research Council Future Fellowship awarded to A. Scheuermann (FT180100692).

\section*{Notations} 
The following symbols are used in this paper:

\begin{longtable}{@{} lp{10cm} @{}}
CF & coarser fraction \\
CL & coarse layer \\
$d_e$ & mean particle diameter \\
$d_{eff}$ & effective particle diameter \\
$d_e^{CF}$ & mean particle diameter of the coarser fraction \\
$d_e^{FF}$ & mean particle diameter of the finer fraction \\
$d_i^{CF}$ & average diameter of particles in the $i^{th}$ bin of the particle size distribution \\
$E_{onset}$ & cumulative energy expended at the onset condition \\
FF & finer fraction \\
$f_c$ & fines content \\
$G_s$ & specific gravity of solids (glass beads) \\
$H_{infil}(t)$ & infiltration height of finer fraction into the coarse layer \\
$H_{initial}$ & initial height of the interface between the mixture and coarse layers \\
$\Dot{H}_{infil}$ & rate of infiltration of finer fraction into the coarse layer \\
$h_{CL}$ & height of coarse layer \\
$h_{ML}$ & height of mixture layer \\
$i_{avg}$ & average hydraulic gradient \\
$i_{avg}^{CL}$ & average hydraulic gradient across the coarse layer \\
$i_{avg}^{ML}$ & average hydraulic gradient across the mixture layer \\
$i_{avg}^{ML+CL}$ & average hydraulic gradient across the test specimen \\
$i_{CM}$ & hydraulic gradient at the complete mixture condition \\
$i_{crit}^{Terz}$ & critical hydraulic gradient for heave failure \\ 
$i_{onset}$ & hydraulic gradient at the limiting onset condition observed in the experiments \\
$i_{onset}^{Darcy}$ & hydraulic gradient in the mixture layer at the limiting onset condition from conventional Darcy approach \\
$i_{onset}^{TDR}$ & hydraulic gradient in the mixture layer at the limiting onset condition from spatial TDR approach \\
$k_{avg}^{Chapuis}$ & average hydraulic conductivity based on the empirical equation proposed by \citet{chapuis2004predicting} \\
$k_{avg}^{Darcy}$ & average hydraulic conductivity based on Darcy’s law \\
ML & mixture layer \\
$m_f$ & mass of finer fraction \\
$\Dot{m}_{cum}^{CL}$ & cumulative gain rate of finer fraction into the coarse layer \\
$\Dot{m}_{cum}^{ML}$ & cumulative loss rate of finer fraction from the mixture layer \\
$m_{f}^{CL}$ & mass of finer fraction migrated into the coarse layer \\
$m_{f}^{ML}$ & residual mass of the finer fraction in the mixture layer \\
$\Dot{m}_{insta}^{CL}$ & instantaneous rate of change in the mass of the finer fraction in the coarse layer \\
$\Dot{m}_{insta}^{ML}$ & instantaneous rate of change in the mass of the finer fraction in the mixture layer \\
$n_{avg}$ & average porosity \\
$n_{avg}^{CL}$ & average porosity of coarse layer \\
$n_{avg}^{ML}$ & average porosity of mixture layer \\
$n_{avg, CF}$ & average porosity of the coarser fraction \\
$n_{avg, CF}^{ML}$ & average porosity of the coarser fraction in the mixture layer \\
$n_{avg, FF}$ & average porosity of the finer fraction \\
$O_{50, CF}^{ML}$ & average pore diameter of the coarser fraction in the mixture layer \\
$P$ & total flow power \\
$\Bar{P}$ & average flow power between onset and complete mixture conditions \\
PSD & particle size distribution \\
$Q$ & flow rate \\
$Q_{CM}$ & flow rate corresponding to the complete mixture condition \\
$Q_{onset}$ & flow rate corresponding to limiting onset condition \\
TDR & time domain reflectometry \\
$t_{CM}$ & time corresponding to complete mixture condition \\
$t_{onset}$ & time corresponding to limiting onset condition \\
$V$ & total volume of the mixture layer \\
$V_{CF}$ & solid volume of the coarser fraction in the mixture layer \\
$V_{FF}$ & solid volume of the finer fraction \\
$V_S$ & total solid volume of the mixture layer \\
$\Delta f_i^{CF}$ & weight of particles in the $i^{th}$ bin of the PSD \\ 
$\Delta h$ & hydraulic head drop across the upstream and downstream sections of the specimen \\
$\alpha$ & stress reduction factor \\
$\alpha_D$ & shape factor \\ 
$\alpha_{Li}$ & stress reduction factor based on the empirical expression proposed by \citet{li2008seepage} \\
$\alpha_{Terz}^{Darcy}$ & stress reduction factor from the conventional Darcy approach \\
$\alpha_{Terz}^{TDR}$ & stress reduction factor from the spatial TDR approach \\
$\gamma_w$ & unit weight of water \\
$\rho_d$ & dry density
\end{longtable}

\bibliographystyle{ms}
\bibliography{ms}

\end{document}